\documentclass[twocolumn, 10pt]{IEEEtran}
\IEEEoverridecommandlockouts
\usepackage{amsfonts}
\usepackage{amssymb}
\usepackage[cmex10]{amsmath}
\usepackage{graphics}
\usepackage{cite}
\usepackage[dvips]{graphicx}

\usepackage{pgf, tikz}
\usetikzlibrary{arrows, automata}
\usepackage{tikz-cd}

\newtheorem{thrm}{\textbf{Theorem}}

\newtheorem{lemma}{\textbf{Lemma}}

\newtheorem{corol}{\textbf{Corollary}}

\newcommand{\abs}[1]{\left\vert#1\right\vert}
\newcommand{\norm}[1]{\Vert#1\Vert}

\ifCLASSINFOpdf
\else
\fi
\begin{document}
%
\title{Time Complexity Analysis of a Distributed Stochastic Optimization in a Non-Stationary Environment}
\author{\IEEEauthorblockN{B. N. Bharath and P. Vaishali\\}
\IEEEauthorblockA{Dept. of ECE, PESIT Bangalore South Campus,\\ Bangalore 560100, INDIA\\}
E-mail: \texttt{bharathbn@pes.edu}, \texttt{vaishali.p.94@gmail.com}
}


%


\maketitle

\begin{abstract}
In this paper, we consider a distributed stochastic optimization
problem where the goal is to minimize the time average of a cost function subject to a set of constraints on the time averages
of related stochastic processes called penalties. We assume that
the state of the system is evolving in an independent and non-stationary
fashion and the ``common information" available at each node is distributed and delayed. 
Such stochastic optimization is an integral part of many important problems in wireless networks such as scheduling, routing, resource allocation and crowd sensing.
We propose an approximate
distributed Drift-Plus-Penalty (DPP) algorithm, and show that
it achieves a time average cost (and penalties) that is within $\epsilon > 0$ of the optimal cost (and constraints) with high probability. Also, we provide a condition on the convergence time $t$ for this result to hold. In particular, for any delay $D\geq 0$ in the common information, we use a coupling argument to prove that the proposed algorithm converges \emph{almost surely} to the optimal solution. We use an application from wireless sensor network to corroborate our theoretical findings through simulation results.




\end{abstract}


%
\IEEEpeerreviewmaketitle

\section{Introduction} \label{sec:intorduction}
Stochastic optimization is ubiquitous in various domains such as communications, signal processing, power grids, inventory control for product assembly systems and dynamic wireless networks \cite{ciftcioglu2013maximizing, neely2013dynamic,chia_energycoop_wcnc2013,neely2016distributed, urgaonkar2011optimal, baghaie2010energy, neely2010dynamic, neelyinfocom2015, xuzhu_shengICC_2014}. A typical stochastic optimization problem involves designing control action for a given state of the system that minimizes the time average of a cost function subject to a set of constraints on the time average penalties \cite{ciftcioglu2013maximizing, neely2013dynamic}. Both cost and penalties depend on the state of the system and the control actions taken by the users. For example, in a typical wireless application, the cost function refers to the instantaneous rate, and the penalty refers to the instantaneous power consumed. Further, the state here refers to the channel conditions. An algorithm known as Drift-Plus-Penalty (DPP) (see \cite{neely2010stochastic,neely2008fairness,georgiadis2006resource, neely2010efficient, samarakoon2015energy}) is known to provide a solution for these problems with theoretical guarantees. At each time slot, the DPP method, an extension of the back-pressure algorithm \cite{tassiulas1992stability,tassiulas1993dynamic}, finds a control action that minimizes a linear combination of the cost and the drift. In the problem that we consider, the drift is a measure of the deviation (of the penalties) from the constraints, and the penalty corresponds to the cost. The DPP algorithm is shown to achieve an approximately optimal solution even when the system evolves in a non-stationary fashion, and is robust to non-ergodic changes in the state \cite{neely2010stochastic}. 

The DPP algorithm mentioned above assumes that the control action is taken at a centralized unit where the complete state information is available. However, wireless network and crowd sensing applications require a distributed control action that uses only the delayed state information at each node \cite{han2014distributed,neely2010stochastic}. This calls for a distributed version of the DPP algorithm with theoretical guarantees. The author in \cite{neely2016distributed} considers a relaxed version of the above problem. In particular, assuming i.i.d. states with correlated ``common information," the author in \cite{neely2016distributed} proposes a distributed DPP algorithm, and proves that the approximate distributed DPP algorithm is close to being optimal. Several authors use the above results in various contexts such as crowd sensing \cite{han2014distributed}, energy efficient scheduling in MIMO systems \cite{zhang_wcnc2013}, 
to name a few. However, in many practical applications, the states evolve in a dependent and non-stationary fashion \cite{neely2010efficient}. Thus, the following assumptions about the state made in \cite{neely2016distributed} need to be relaxed: (i) independent and (ii) identically distributed. Further, from practical and theoretical standpoints, it is important to investigate the rate of convergence of the distributed algorithm to the optimal. In this paper, we relax the assumption (ii) above, and unlike \cite{neely2016distributed}, we provide a Probably Approximately Correct (PAC) bound on the performance. Also, we prove an \emph{almost sure} convergence of the proposed \emph{distributed} algorithm to a constant within the optimal. We would like to emphasize that extending the analysis in \cite{neely2016distributed} to non-stationary states is non-trivial. 
The only work that provides a ``PAC type" result for the DPP algorithm is \cite{wei2015sample}. However, the authors consider i.i.d. states, and the decision is \emph{centralized}. Moreover, the method used in \cite{wei2015sample} cannot be directly extended to a problem with non-stationary states since their proof requires the control action to be stationary, and this assumption in general is not true. Now, we highlight the contribution of our work.

\subsection{Main Contribution of the Paper}\label{subsec:main_contributions}
In this paper, we consider a distributed stochastic optimization problem when the states evolve in an independent and \emph{non-stationary} fashion. In particular, we assume that the state is asymptotically stationary, i.e., the probability measure $\pi_t$ of the state $\omega(t) \in \Omega$ converges to a probability measure $\pi$ as $t \rightarrow \infty$ in the $\mathcal{L}_1$-norm sense. This assumption makes the extension of the method in \cite{neely2016distributed} non-trivial. When $\pi_t = \pi$ for all $t \in \mathbb{N}$, the author in \cite{neely2016distributed} proves theoretical guarantees by making use of the equivalence between a Linear Program (\textbf{LP}) that is a function of $\pi$ and the original stochastic optimization problem. However, when the probabilities are changing, this equivalence is difficult to establish. Instead, we show that the original problem is equivalent to a ``perturbed" \textbf{LP}, which is a function of the limiting distribution $\pi$. Under mild conditions, we prove that the solution to the perturbed \textbf{LP} is approximately equal to that of the original problem. We use this result to prove theoretical guarantees for an approximate DPP algorithm that we propose in the paper. Moreover, unlike the previous works, we are more interested in providing sample complexity bounds rather than just dealing with the averages. The following are the main contributions of our work 

\begin{enumerate}[]
\item For the above model, we show that with high probability, the average cost and penalties obtained by using the proposed approximate \emph{distributed} DPP are within constants of the optimal solution and the constraints, respectively, provided the waiting time $t > \text{a threshold}$ (see Theorem \ref{thm:mainresult1}). The threshold and the constants capture the degree of non-stationarity (i.e., $\norm{\pi_t - \pi}_1$), and the number of samples used to compute an estimate of the state distribution. 
\item Using the high probability result, we show that the cost corresponding to the proposed algorithm \emph{almost surely} converges to a constant within $\epsilon_0>0$ of the optimal cost. We also show that the penalties induced by the proposed algorithm are within constants of the constraint values \emph{almost surely}.
It turns out that although the states are independent, the proposed algorithm induces dependencies across time in the cost and penalties. To prove the PAC and the almost sure convergence results, we use a coupling argument where the dependent sequence of the cost (also, penalties) is replaced by an independent sequence which results in an error expressed in terms of the $\beta_1$-mixing coefficient; a term that captures the stochastic dependency across time (see Sec. \ref{sec:sys_model}). The $\beta_1$-mixing coefficient is bounded using information theoretic techniques to complete the proof. 
\item We show that due to non-stationarity of the states, the performance gap goes down slowly compared to i.i.d. states. This is captured through $\norm{\pi_t - \pi}_1$ and a term that depends on the measure of the complexity of  the probability space averaged with respect to $\pi_t$ (see Theorem \ref{thm:mainresult1}). Finally, we provide simulation results of a sensor network application, which is a particular use case scenario of the problem considered.
\end{enumerate}
The paper is organized as follows. The problem statement, an approximate DPP Algorithm with related theoretical guarantees and simulation results are provided in  Sec. \ref{sec:sys_model}, Sec. \ref{sec:alg_result} and Sec. \ref{sec:sims}, respectively. A bound on the mixing coefficient is provided in Sec. \ref{sec:mixing_bound}. Sec. \ref{sec:conl} concludes the paper.  \\
\textbf{Notation:} We use the following notations in the paper. We write $f(x) \doteq g(x)$, $f(x) \preceq g(x)$, $f(x) \prec g(x)$, $f(x) \succeq g(x)$, and $f(x) \succ g(x)$ to mean $\lim_{x \rightarrow \infty}\frac{f(x)}{g(x)} = 1$, $\lim_{x \rightarrow \infty}\frac{f(x)}{g(x)} \leq 1$, $\lim_{x \rightarrow \infty}\frac{f(x)}{g(x)} < 1$, $\lim_{x \rightarrow \infty}\frac{f(x)}{g(x)} \geq 1$ and $\lim_{x \rightarrow \infty}\frac{f(x)}{g(x)} > 1$, respectively. We use $f(x) = \mathcal{O}(g(x))$ if $\lim_{x \rightarrow \infty}\frac{f(x)}{g(x)} = c$ for some $c < \infty$.

\section{Motivation and Problem Statement} \label{sec:sys_model}
Towards motivating the system model studied in the paper, we consider a network of $3$ sensors, where the sensor $i$ observes the state $\omega_i(t) \in \{0,1,2,3\}$, $i=1,2,3$, and reports the observation to a central unit \cite{neely2016distributed}. The reporting incurs a penalty in terms of the power consumed by the sensors to transmit the state information. The state $\mathbf{\omega}{(t)}\triangleq\{\omega_1(t),\omega_2(t), \omega_3(t)\}$, $t \in \mathbb{N}$ in general is a stochastic process that evolves in a \emph{non-stationary} fashion. Assume that the central unit trusts sensor $1$ more than the others. The problem is to maximize the average of the following utility function subject to the constraint that the average power consumed by each sensor is less than $\bar{P}$: 
\begin{equation} \label{eq:example}
u_0(t)\triangleq \min\left\{\frac{\alpha_1(t) \omega_1(t)}{3} + \frac{\alpha_2(t) \omega_2(t) + \alpha_3(t) \omega_3(t)}{6},1\right\},
\end{equation}
where $\alpha_i(t) \in \{0,1\}$, $i=1,2,3$ are the decision variables. Note that if $\omega_i(t) = 3$ for $i=1,2,3$, and $\alpha_i(t) = 1$ for $i=2,3$, then there is no increase in the utility if sensor $1$ also decides to transmit, i.e., $\alpha_1(t) = 1$. However, none of the sensors know the entire state of the system. In this case, the sensor $1$ may also choose to transmit, thus wasting its power leading to a suboptimal operation compared to a centralized scheme. In order to resolve this issue in a distributed setting, we assume that a delayed ``common information" is available (see Sec. II of \cite{neely2016distributed} for more details) using which each sensor picks one of the ``pure strategies". For example, each sensor can acquire the information about the state $\omega(t)$ with a fixed delay $D>0$. In this case, the ``common information" can be some function of $\omega(t-D)$. Thus, the problem is to find the set of optimal decision variables in a distributed fashion with ``common information" that maximizes the average of the above utility subject to the constraints on the average power. Next, we describe the system model that generalizes the above example, and later provide an algorithm with theoretical guarantees. 

Consider a system comprising of $N$ users making decisions in a distributed fashion at discrete time steps $t\in \{0,1,2,\ldots\}$. Each user $i$ observes a random state $\omega_i(t) \in \Omega_i$, and a ``common information" $Y_c(t) \in \mathcal{Y}$ to make a control decision $\alpha_i(t) \in \mathcal{A}_i$, $i=1,2,\ldots, N$. Here, for each user $i$, $\Omega_i$, $\mathcal{Y}$ and $\mathcal{A}_i$ denote the state space, common information space and action/control space, respectively.  Let ${\omega}(t) \triangleq \{\omega_1(t),\omega_2(t),\ldots,\omega_N(t)\} \in \Omega$ and $\alpha(t)\triangleq\{\alpha_1(t),\alpha_2(t),\ldots,\alpha_N(t)\} \in \mathcal{A}$, where $\Omega \triangleq \Omega_1 \times \Omega_2 \times \ldots \times \Omega_N$, and $\mathcal{A} \triangleq \mathcal{A}_1 \times \mathcal{A}_2 \times \ldots \times \mathcal{A}_N$. Also, let us assume that the number of possible values that $p_k(t)$ takes is finite and equal to $\mu_k \in \mathbb{N}$, $k=1,\ldots,K$. The decision is said to be \emph{distributed} if (see \cite{neely2016distributed})
\begin{itemize}
\item There exists a function $f_i : \Omega_i \times \mathcal{Y} \rightarrow \mathcal{A}_i$, such that 
\begin{equation} \label{eq:dist_condition}
\alpha_i(t)\triangleq f_i(\omega_i(t), Y_c(t)),~ i=1,2,\ldots, N,
\end{equation}
where $Y_c(t)$ belongs to the common information set $\mathcal{Y}$.
\item The common information $Y_c(t)$ is independent of $\omega(t)$ for every $t \in \mathbb{N}$.
\end{itemize}

At each time slot $t$, the decision $\alpha(t)$ and the state $\omega(t)$ result in a cost $p_0(t)\triangleq p_0(\alpha(t),\omega(t))$ and penalties $p_k(t)\triangleq p_k(\alpha(t),\omega(t))$, $k=1,2,\ldots,K$. The central goal of the paper is to analyze an approximate distributed solution to the following problem when $\omega(t)$, $t\in \mathbb{N}$ is independent and \emph{non-stationary}, $\mathbf{P_0:}$
\begin{eqnarray}
& \min_{\alpha(\tau) \in \mathcal{A}: \tau \in \mathbb{N}}& \limsup_{t \rightarrow \infty} \frac{1}{t}\sum_{\tau=0}^{t-1} \mathbb{E}p_0({\tau}) \nonumber \\
\hspace{-1.6cm} &\text{subject to}&\hspace{-0.5cm} \limsup_{t \rightarrow \infty} \frac{1}{t}\sum_{\tau=0}^{t-1} \mathbb{E}p_k({\tau}) \leq c_k,~k=1,2,\ldots, K, \nonumber \\
&& \hspace{-0.8cm} \alpha_i(\tau) \text{ satisfies \eqref{eq:dist_condition}, } i=1,2,\ldots,N. \nonumber
\end{eqnarray} 
In the above, the expectation is jointly with respect to the distribution of the state $\omega(t)$ and a possible randomness in the  decision $\alpha(t)$, $t \in \mathbb{N}$. Let $p^{(opt)}$ be the optimal cost corresponding to the problem $\mathbf{P_0}$. Note that the first equation in $\mathbf{P_0}$ represents the time average cost while the second and the third equations represent constraints on the penalties and the decisions, respectively. Informally, we are interested in proving a Probably Approximately Correct (PAC) type result of the following form \cite{wei2015sample}
\begin{itemize}
\item For every $\epsilon_k > 0$, with a probability of at least $1 - \delta_k$, $\frac{1}{t} \sum_{\tau = 0}^{t-1} p_k^{(\approx)}(\tau) \leq c_k + \epsilon_k$ provided $t > $ a threshold, where $p_0^{(\approx)}(\tau)$ and $p_k^{(\approx)}(\tau)$, $k=1,2,\ldots,K$ are the cost and penalties, respectively, of an approximate distributed scheme at $\tau \in \mathbb{N}$. Here $c_0 \triangleq p^{(\textit{opt})}$ is the optimal cost, and $c_k$, $k=1,2,\ldots,K$ are as defined in $\mathbf{P_0}$.
\end{itemize}
First, unlike the model in \cite{neely2016distributed}, we assume that the state $\omega(t)$ evolves in an independent and \emph{non-stationary} fashion across time $t$. In particular, the distribution of $\omega(t)$ denoted $\pi_t(\omega)$, $\omega \in \Omega$ satisfies the following asymptotic stationarity property.

\textbf{Assumption 1:} Assume that there exists a probability measure $\pi(\omega)$ on $\Omega$ such that $$\lim_{t \rightarrow \infty} \norm{\pi_t - \pi}_{1} = 0.$$

Note that the efficacy of the distributed algorithm depends on how accurately each node computes an estimate of $\pi_t$, $t \in \mathbb{N}$. Naturally, we expect the bounds that we derive to be a function of the complexity of the probability measure space from which the ``nature" chooses $\pi_t(\omega)$. Let us assume that for each $t \in \mathbb{N}$, $\pi_t$ is chosen from a set $\mathcal{P}$. Assuming that $\mathcal{P}$ is a closed set with respect to the $\mathcal{L}_1$\emph{-norm}, we have $\pi \in \mathcal{P}$. One way of measuring the complexity is through the covering number, and the metric entropy of the set $\mathcal{P}$, which are defined as follows.

\textbf{Definition 1: }(see \cite{van2000applications}) A $\delta$-covering of $\mathcal{P}$ is a set $\mathcal{P}_c \triangleq \{\mathcal{P}_1,\mathcal{P}_2,\ldots,\mathcal{P}_M\} \subseteq \mathcal{P}$ such that for all $\pi^{'} \in \mathcal{P}$, there exists a $\mathcal{P}_i \in \mathcal{P}_c$ for some $i=1,2,\ldots,M$ such that $\norm{\pi^{'} - \mathcal{P}_i}_1 < \delta$. The smallest $M$ denoted $M_{\delta}$ is called the covering number of $\mathcal{P}$. Further, $\mathcal{H}(\mathcal{P},\delta) \triangleq \log M_\delta$ is called the \emph{metric entropy}.


Note that in many practical scenarios, the available data at each time $t \in \mathbb{N}$ is delayed, and a data of size $w_t$, $t \in \mathbb{N}$ delayed by $D$ slots will be used for estimation/inference purposes \cite{neely2016distributed, han2014distributed}. The reason for making the sample size $w_t$ depend on $t$ becomes apparent later. 
Since $p_k(t)$, $k=0,1,2,\ldots,K$ depend on $Y_c(t)$ for all $t$ (see \eqref{eq:dist_condition}), we have that the  process $p_k(t)$ in general is a stochastically dependent sequence. The ``degree" of correlation depends on the algorithm used. For $k=0,1,2\ldots,K$ and $s \in \mathbb{N}$, let $\mathbb{P}^{\texttt{ALG},k}_{t, t+s}(* \left \vert \right. \mathcal{E})$ and $\mathbb{P}^{\texttt{ALG},k}_{t}(*\left \vert \right. \mathcal{E})$ denote the joint and marginal distributions of $(p_k(t), p_k(t+s))$ and $p_k(t)$ conditioned on the event $\mathcal{E}$, respectively, induced by any algorithm $\texttt{ALG}$.\footnote{In this paper, we propose a distributed Approximate DPP (ADPP) algorithm, and hence \texttt{ALG} will be \texttt{ADPP}.} Note that if $p_k(t)$ and $p_k(t+s)$ are independent for each $t \in \mathbb{N}$ conditioned on some event $\mathcal{E}$, then $\left \vert \left \vert \mathbb{P}^{\texttt{ALG},k}_{t, t+s}(* \left \vert \right. \mathcal{E}) -  \mathbb{P}^{\texttt{ALG},k}_{t}(*\left \vert \right. \mathcal{E}) {\otimes}  \mathbb{P}^{\texttt{ALG},k}_{t + s}(*\left \vert \right. \mathcal{E})\right \vert \right \vert_{\texttt{TV}} = 0$. Thus, the difference above, maximized over all slots $t \in \mathbb{N}$ is a natural way of measuring the correlation between the sequences that are $s$ time slots away. More precisely, we have the following definition (see \cite{kuznetsov2014generalization} for a related definition).\\
\textbf{Definition 2:} The $\beta_1$ mixing coefficient of the process $p_k(t)$, $k=0,1,2,\ldots,K$ conditioned on some event $\mathcal{E}$ is given by
\begin{equation}
\beta_{\texttt{ALG},k}(s,\alpha \left \vert \right. \mathcal{E}) \triangleq \sup_{t \in \mathbb{N}, t \geq \alpha} \norm{\mathbb{M}_{t,s,k}(\mathcal{E})}_\texttt{TV},
\end{equation}
where $\mathbb{M}_{t,s,k}(\mathcal{E}) \triangleq \mathbb{P}^{\texttt{ALG},k}_{t, t+s}(*\left \vert \right. \mathcal{E}) - \mathbb{P}^{\texttt{ALG},k}_{t}(*\left \vert \right. \mathcal{E}) \otimes \mathbb{P}^{\texttt{ALG},k}_{t + s}(*\left \vert \right. \mathcal{E})$, $s \geq 0$, $\alpha \geq 0$, $\mathbb{P}^{\texttt{ALG},k}_{t} \otimes \mathbb{P}^{\texttt{ALG},k}_{t + s}$ denotes the product distribution, and $\norm{*}_\texttt{TV}$ is the total variational norm.

Note that in the definition of $\beta_{\texttt{ALG},k}(s,\alpha \left \vert \right. \mathcal{E})$, we have used $t \geq \alpha$, which is required later in the proof of our main results. Further, if $s$ is large, and the process is sufficiently mixing, then we expect that $\beta_{\texttt{ALG},k}(s,\alpha \left \vert \right. \mathcal{E}) = 0$. This definition will be used to decouple a dependent stochastic process so that some of the large deviation bounds that are valid for independent sequences can be applied. The details of this approach will be clear in the proof of our main results. For notational convenience, let us denote the maximum and minimum values of $p_k(t)$, $k=0,1,2,\ldots,K$ by $p_{\text{max},k}$ and $p_{\text{min},k}$, respectively. Further, let $(\Delta p)_{\text{max},k} \triangleq p_{\text{max},k} - p_{\text{min},k}$. In the following section, we propose an Approximate DPP (ADPP) algorithm with the associated theoretical guarantees. The $\beta_1$ coefficient for the ADPP algorithm will be $\beta_{\texttt{ADPP},k}(s,\alpha \left \vert \right. \mathcal{E})$.  

\section{Algorithm and Main Results} \label{sec:alg_result}
In the following subsection, we prove that the optimal solution to $\mathbf{P_0}$ is close to a \textbf{LP}. 
\subsection{Approximately Optimal LP}
Since the number of possible values that $p_k(t)$, $k=0,1,2,\ldots,K$ take is finite, the number of possible strategies is also finite.\footnote{Due to this, the question of whether $p_k(t)$ is convex or not does not matter.} The approximate algorithm that we are going to propose chooses one of the \emph{pure strategy} $\mathbf{S(\omega)} \triangleq \{\mathbf{s}_1(\omega_1), \mathbf{s}_2(\omega_2), \ldots, \mathbf{s}_N(\omega_N)\}$ based on the common information $Y_c(t)$,
where $\mathbf{s}_i(\omega_i) \in \mathcal{A}_i$, and $\omega_i \in \Omega_i$, $i=1,2,\ldots,N$. For example, $\mathbf{s}_i(\omega_i)$ can be a simple threshold rule with the thresholds coming from a finite set. The control action $\alpha_i(t)$ at the user $i$ is chosen as a deterministic function of $\omega(t)$, i.e., $\alpha_i(t)\triangleq \mathbf{s}_i(\omega_i(t))$ for all $i \in \{1,2,\ldots, N\}$ and for all $t \in \mathbb{N}$. Let the total number of such pure strategies be $F\triangleq \prod_{i=1}^N \abs{\mathcal{A}_i}^{\abs{\Omega_i}}$. Enumerating the $F$ strategies, we get $\mathbf{S}^m(\omega)$, $m \in \{1,2,\ldots,F\}$ and $\omega \in \Omega$. Each $\omega \in \Omega$ and the strategy $\mathbf{S}^m(\omega)$ result in a cost $p_k(\mathbf{S}^m(\omega), \omega)$, $k=0,1,2,\ldots,K$. Note that it is possible to reduce $F$ if the problem has a specific structure  \cite{neely2016distributed}. For each strategy $m \in \{1,2,\ldots,F\}$, define the average cost/penalty as 
\begin{eqnarray}
r_{k,\pi^{'}}^{(m)} &\triangleq& \sum_{\omega \in \Omega} \pi^{'}(\omega) p_k(\mathbf{S}^{m}(\omega), \omega), \label{eq:rkpit}
\end{eqnarray}
where $k = 0,1,2,\ldots,K$ and the underlying distribution of $\omega \in \Omega$ is $\pi^{'} \in \mathcal{P}_c$. 
As in \cite{neely2016distributed}, we consider a randomized algorithm where the strategy $m \in \{1,2,\ldots,F\}$ is picked with probability $\theta_m(t)$ in an independent fashion across time $t$. Here, $\theta_m(t)$ is a function of the common information $Y_c(t)$. The corresponding average cost/penalty at time $t$ becomes
\begin{eqnarray}
\mathbb{E} p^{}_k(t) &=& \sum_{m=1}^F \theta_m(t) \mathbb{E}_{\lambda} p_k({\mathbf{S}^{m}}(\omega(t)), \omega(t)) \nonumber \\ 
&=& \sum_{m=1}^F \theta_m(t) r_{k,\lambda}^{(m)},\nonumber
\end{eqnarray}
where $\lambda \in \{\pi_t, \pi, \mathcal{P}_i\}$, $i=1,2,\ldots, M_\delta$. In \cite{neely2016distributed}, it was shown that the problem $\mathbf{P_0}$ when $\pi_t = \pi$ for all $t \in \mathbb{N}$ ($\omega(t)$ is i.i.d.) is equivalent to the following \textbf{LP}:
\begin{eqnarray} \label{eq:LPneely}
&\min_{\theta_1,\theta_2,\ldots,\theta_F}& \sum_{m=1}^F \theta_m r_{0,\pi}^{(m)} \nonumber\\
&\text{subject to}& \sum_{m=1}^F \theta_m r_{k,\pi}^{(m)} \leq c_k, ~k=1,2,\ldots, K \nonumber\\
&&\sum_{m=1}^F \theta_m = 1. \label{eq:lp}
\end{eqnarray}
In this paper, from \textbf{Assumption 1}, we have $\norm{\pi_t - \pi}_1 \rightarrow 0$, as $t \rightarrow \infty$. With dense covering of the space $\mathcal{P}$, we expect that the limiting distribution is well approximated by $\mathcal{P}_i$ for some $i=1,2,\ldots,M_\delta$ in the covering set. More preciesely, $$\mathcal{P}_{i^{*}} \triangleq \arg \min_{\mathcal{Q} \in \{\mathcal{P}_1,\ldots,\mathcal{P}_{M_\delta}\}} \norm{\pi - \mathcal{\mathcal{Q}}}_1,$$ and the corresponding distance be $d_{\pi,\mathcal{P}_{i^{*}}} \triangleq \norm{\pi - \mathcal{P}_{i^{*}}}_1 < \delta$.  Since the distribution of $\omega(t)$ is changing across time, directly applying Theorem $1$ of \cite{neely2016distributed} is not possible. However, from \textbf{Assumption $1$}, we know that the distribution approaches a fixed measure $\pi \in \mathcal{P}_c$. Hence, we expect that the algorithm designed for $\pi \in \mathcal{P}_c$ or an approximation of $\pi$, i.e., $\mathcal{P}_{i^{*}}$ should eventually be close to the optimal algorithm. Therefore, we consider the following \textbf{LP} denoted $\mathbf{LP_{\mathcal{P}_{i^{*}}}}$:
\begin{eqnarray} \label{eq:LP2}
&\min_{\theta_1,\theta_2,\ldots,\theta_F}& \sum_{m=1}^F \theta_m r_{0,\mathcal{P}_{i^{*}}}^{(m)} \nonumber\\
&\text{subject to}& \sum_{m=1}^F \theta_m r_{k,\mathcal{P}_{i^{*}}}^{(m)} \leq c_k, ~k=1,2,\ldots, K \nonumber\\
&&\sum_{m=1}^F \theta_m = 1. \label{eq:lp2}
\end{eqnarray}
Also, we assume that the solution to $\mathbf{LP_{\mathcal{P}_{i^{*}}}}$ exists and the optimal cost is absolutely bounded. Further, define
\begin{equation} \label{eq:gxlp}
G(x) \triangleq \inf\left\{\sum_{m=1}^F \theta_m r_{0,\mathcal{P}_{i^{*}}}^{(m)}: \Theta \in \mathcal{C}_{x,\Theta}\right\},
\end{equation}
where $\Theta \triangleq (\theta_1,\theta_2,\ldots, \theta_F)$, and for any $x \geq 0$, 
\begin{eqnarray}
&&\hspace{-0.8cm}\mathcal{C}_{x,\Theta} \triangleq \nonumber \\
&&\hspace{-0.8cm}\left\{\Theta: \sum_{m=1}^F \theta_m r_{k,\mathcal{P}_{i^{*}}}^{(m)} \leq c_k + x, ~k=1,2,\ldots, K, \Theta \mathbf{1}^T = 1\right\} \nonumber,
\end{eqnarray}
where $\mathbf{1} \triangleq \{1,1,\ldots,1\} \in \mathbb{R}^F$. Note that $G(0)$ corresponds to $\mathbf{LP_{\mathcal{P}_{i^{*}}}}$. We make the following important smoothness assumption about the function $G(x)$. 

\noindent \textbf{Assumption 2:} The function $G(x)$ is \emph{$c$-Lipschitz continuous} around the origin, i.e., for some $c > 0$, we have
\begin{equation}
\abs{G(x) - G(y)} \leq c \abs{x-y}, \text{ for all } x,y \geq 0.
\end{equation}

In the theorem to follow, given that \textbf{Assumption 2} is valid, we prove that the optimal cost of the linear optimization problem in \eqref{eq:LP2} is ``close" to the optimal cost of $\mathbf{P_0}$.

\begin{thrm} \label{thm:lppit_popt_relation}
Let $p^{\text{(opt)}}$ and $p^{(\text{opt})}_{\mathcal{P}_{i^{*}}}$ be the optimal solution to the problems $\mathbf{P_0}$ and $\mathbf{{LP}_{\mathcal{P}_{i^{*}}}}$, respectively. Then, under \textbf{Assumption 2}, we have $p^{(\text{opt})}_{\mathcal{P}_{i^{*}}} < p^{\text{(opt)}} + (c + 1) \Delta_{\pi,\mathcal{P}_{i^{*}}}$, where for any $\nu >0$, $\Delta_{\pi,\mathcal{P}_{i^{*}}} \triangleq \max_{k=0,1,2,\ldots,K} b_{\text{max},k}  (d_{\pi,\mathcal{P}_{i^{*}}} + \nu), \text{ and } b_{\text{max},k} \triangleq \max\{\abs{p_{\text{max},k}}, \abs{p_{\text{min},k}}\}.$ 
\end{thrm}
\emph{Proof:} See Appendix \ref{app:lppit_popt_relation}. $\blacksquare$


\subsection{Approximate DPP (ADPP) Algorithm} \label{subsec:dpp}
In this subsection, we present an online distributed algorithm that approximately solves the problem $\mathbf{P_0}$. We assume that at time $t \in \mathbb{N}$, all nodes receive feedback specifying the values of all the penalties and the states, namely, $p_1(t-D), p_2(t-D), \ldots, p_K(t-D)$ and $\omega(t-D)$. Recall that $D \geq 0$ is the delay in the feedback. Using this information, we construct the following set of queues 
\begin{equation} \label{eq:queue_update}
Q_k(t+1) = \max\{Q_k(t) + p_k(t-D) - c_k, 0\},
\end{equation}
$k=1,2,\ldots,K$, and $t \in \mathbb{N}$. These queues act as the common information, i.e., $Y_c(t) = \mathbf{Q}_t$, where $\mathbf{Q}_t \triangleq (Q_1(t),Q_2(t),\ldots,Q_K(t))$. Further, the past $w_t$ samples of $\omega(t)$ given by $\{\omega(t-i)$, $i=D,D+1,\ldots,D+w_t-1\}$ will be used to find an estimate of the state probabilities which is required for the algorithm that we propose. For all $k=1,2,\ldots, K$, we let $p_k(t) = 0$ when $t \in \{-1,-2,\ldots, -D\}$. The \emph{Lyapunov} function is defined as 
\begin{equation}
\mathcal{L}(t) \triangleq \frac{1}{2} \norm{\mathbf{Q}_t}^2_2 = \frac{1}{2} \sum_{i=1}^K Q_i^2(t),
\end{equation}
and the corresponding drift is given by $\Delta(t)\triangleq\mathcal{L}(t+1) - \mathcal{L}(t)$ for all $t \in \mathbb{N}$. A higher value of the drift indicates that the constraints have been violated frequently in the past. Thus, the control action should be taken that simultaneously minimizes the drift and the penalty (cost). The DPP algorithm tries to find the optimal control action that minimizes an upper bound on the DPP term $\mathbb{E} \left[\Delta(t+D) + V p_0(t) \left \vert \right. \mathbf{Q}_t\right]$, $V \geq 0$, which is the essence of the following lemma. The proof of the lemma follows  directly from the proof of Lemma $5$ of \cite{neely2016distributed}, and hence omitted.  

\begin{lemma} 
For a fixed constant $V \geq 0$, we have
\begin{eqnarray} \label{eq:dpp_bound_lemma}
&&\mathbb{E} \left[\Delta(t+D) + V p_0(t) \left \vert \right. \mathbf{Q}_t\right] \leq B_t(1 + 2D) +\nonumber \\ 
&& V \sum_{m=1}^F \beta_m(t) r_{0,\pi_t}^{(m)} + \sum_{k=1}^K Q_k(t) \mathcal{C}_{i,k,t},
\end{eqnarray}
where $\mathcal{C}_{i,k,t} \triangleq  \sum_{m=1}^F \beta_m(t) r_{k,\pi_t}^{(m)} - c_k $, $r_{k,\pi_t}^{(m)} \triangleq \sum_{\omega \in \Omega} \pi_t(\omega) p_k({\mathbf{S}^{m}}(\omega), \omega)$, $k=0,1,2,\ldots,K$,  
\begin{equation} \label{eq:btau}
~B_t\triangleq \max_{m \in \{1,2,\ldots, F\}} \frac{1}{2} \sum_{k=1}^K \sum_{\omega \in \Omega} \pi_t(\omega) \abs{p_k({\mathbf{S}^{m}}(\omega), \omega) - c_k}^2,
\end{equation}
and, with a slight abuse of notation, $\beta_m(t)$ is the probability with which the strategy $m$ is used at time $t$.
\end{lemma}

Note that as $t \rightarrow \infty$, $B_t \rightarrow B$. The expression for $B$ can be obtained by replacing $\pi_t(\omega)$ by $\pi(\omega)$ in the expression for $B_t$. The algorithm to follow requires an estimate of $\pi_t(\omega)$, which can be computed using the past $w_t$ samples by means of any estimate such as the sample average. However, when the space $\mathcal{P}$ is ``simple", one can expect to compute an estimate of $\pi_t(\omega)$ more efficiently. For example, if the nature chooses $\omega(t)$ from a finite set of distributions ($M_{\delta} < \infty$ for all $\delta > 0$), then estimating the distribution corresponds to a hypothesis testing problem. Hence, by approximating the measure space $\mathcal{P}$ by a finite set of measures $\mathcal{P}_c$ gives us the flexibility to run a hypothesis testing to find an approximate distribution based on the available $w_t$ samples through a likelihood ratio test. In the following, we provide the algorithm. 

\begin{itemize}
\item \textbf{Algorithm:} Given the delayed feedback of size $w_t$ at time slot $t \in \mathbb{N}$, i.e., $\omega(t-i-D)$, and $p_k(t-D)$, $i=0,1,\ldots,w_t-1$ and for $k=1,2,\ldots,K$, perform the following steps
\begin{itemize}
\item \textbf{Step 1:} Find the probability measure in $\mathcal{P}_c$ that best fits the data, i.e., pick $\mathcal{P}_{j_t^*} \in \mathcal{P}_c$ such that 
\begin{equation} \label{eq:detect_covering_alg}
\hspace{-1cm}j_t^* \triangleq \arg \max_{j \in \{1,2,\ldots,M_\delta\}} \frac{1}{w_t}\sum_{\tau=t-D-w_t+1}^{t-D} \log \left({\mathcal{P}_{j}(\omega(\tau))}\right).
\end{equation}
\item \textbf{Step 2:} Choose $m_t \in \{1,2,\ldots, F\}$ (breaking ties arbitrarily) that minimizes the following:
\begin{equation}
V r_{0,\mathcal{P}_{j_t^*}}^{(m_t)} + \sum_{k=1}^K Q_k(t) r_{k,\mathcal{P}_{j_t^*}}^{(m_t)}. 
\end{equation}
\item \textbf{Step 3:} Set $t \rightarrow t + 1$, receive the delayed feedback,  
update the queues using \eqref{eq:queue_update}, and go to \textbf{Step 1}.
\end{itemize}
\end{itemize}
We say that there is an error in the outcome of step $1$ of the algorithm if $\mathcal{P}_{j_t^*} \neq \mathcal{P}_{i^*}$. Recall that $i^*$ corresponds to the index of the probability measure in the covering set that is close to $\pi$ in the $\mathcal{L}_1$ norm sense. The  error event $\mathcal{E}_{\delta, t}$, $t \in \mathbb{N}$ is defined as those outcomes for which $j_t^* \neq i^*$. Further, let $\mathcal{E}_{[\tau:\tau+s]} \triangleq \bigcup_{t=\tau}^{\tau + s} \mathcal{E}_{\delta, t}$ to denote that there is an error in at least one of the time slot in the interval $\tau$ to $\tau+s$. In the following theorem, we state and prove our first result that will be used to prove the PAC type bound for the ADPP algorithm. 

\begin{figure}[h]
	\centering
	\includegraphics[width=8.0cm, height = 3cm]{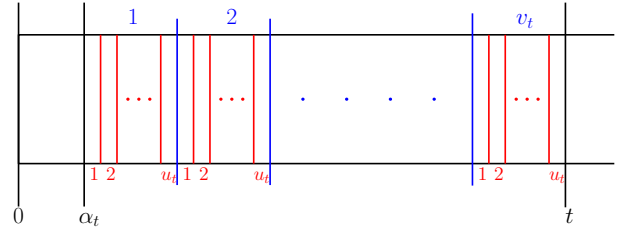}
	\caption{The figure shows the time slot $t - \alpha_t$ split into $v_t$ blocks of size $u_t$ each, i.e., $t - \alpha_t = u_t v_t$. By choosing $\alpha_t = \mathcal{O}(\sqrt t)$, $\mathcal{O}(\sqrt t)$ samples are available at $\tau = \alpha_t$.}
	\label{fig:time_div}
\end{figure}

\begin{thrm} \label{thm:pacfirstresult}
For the $\texttt{ADPP}$ algorithm, for any $\epsilon_k > \frac{1}{t}\sum_{\tau=0}^{t-1} \mathbb{E}p_k(\tau) - c_k + \frac{{\alpha_t}{(p_{\text{max},k} -  p_{\text{min},k})}}{t-\alpha_t}$, and for constants $\alpha_t \in \mathbb{N}$, $u_t \in \mathbb{N}$ and $v_t \in \mathbb{N}$ such that $v_t u_t = t - \alpha_t$, we have 
\begin{eqnarray} \label{eq:mcdiarmid_pac1}
\Pr\left\{\frac{1}{t}\sum_{\tau=0}^{t-1} p_k(\tau) - c_k >  \epsilon_k \right\} &\leq& u_t \exp\left\{\frac{-2 \bar \epsilon_{t,k}^2 v_t^2}{((\Delta p)_{\text{max},k})^2}\right\}+\nonumber \\ 
&& \hspace{-4.5cm}\sum_{\tau = \alpha_t}^t \Pr\left\{\mathcal{E}_{\delta,\tau}\right\}+ (t-\alpha_{t}) \beta_{\texttt{ADPP},k}(u_t, \alpha_t \left \vert \right. \mathcal{E}_{[\alpha_t:t]}^c),
\end{eqnarray}
where $\bar \epsilon_{t,k} \triangleq \frac{t \epsilon_{t,k} - {\alpha_t}(p_{\text{max},k} -  p_{\text{min},k})}{t-\alpha_t}$, $\epsilon_{t,k}\triangleq \epsilon_k + c_k - \frac{1}{t}\sum_{\tau=0}^{t-1} \mathbb{E}p_k(\tau)$. Here, $c_0 = p^{(opt)}$, and $c_k$, $k=1,2,\ldots,K$ are the constraint variables in $\mathbf{P_0}$. 
\end{thrm}
\emph{Proof:} See Appendix \ref{app:pacfirstresult}. $\blacksquare$

The first term in the bound in Theorem \ref{thm:pacfirstresult} corresponds to the large deviation bound when $p_k(t)$'s are independent. The second term corresponds to an upper bound on the probability of error in the time slots $\alpha_t$ to $t$ for decoding the correct index $i^*$; equivalently, this corresponds to an ``incorrect" estimate of the distribution of the states in these slots. The last term captures the stochastic dependency of $p_k(t)$ across time $t \in \mathbb{N}$. In order to prove a high probability result, we need to find an expression for each of the terms in the bound. Next, we upper bound the error term $\Pr\{\mathcal{E}_{\delta, \tau}\}$ using the following assumption about the probability space $\mathcal{P}_c$. This will come handy in the proof of Lemma \ref{lemm:prob_error} below.

\textbf{Assumption 3:} Assume that for all $j=1,2,\ldots,M_\delta$, $\mathcal{P}_{j}(\omega) \neq 0$, there exist constants $\alpha_\delta >\beta_\delta > 0$, such that $\alpha_\delta > \mathcal{P}_{j}(\omega) > \beta_\delta > 0$ for all $\omega \in \Omega$.

We use the above assumption in the proof of the following lemma to bound the probability of error term in \eqref{eq:mcdiarmid_pac1}.
\begin{lemma} \label{lemm:prob_error}
An upper bound on the probability of error is given by 
\begin{equation} \label{eq:peupperbound_stationary_lemma}
 \Pr\{\mathcal{E}_{\delta,\tau}\} \leq  P^{(\tau)}_{e,\texttt{up}} \triangleq \left\{ \begin{array}{cc} 
 q^{(\tau)}_{e,\texttt{up}} & \text{ if } \tau > D + w_\tau - 1,\\
 \frac{1}{M_\delta} & \text{otherwise},
\end{array}
\right.
\end{equation}
where $q^{(\tau)}_{e,\texttt{up}} \triangleq \exp\left\{- {2 \zeta_\delta \mathcal{D}_\tau^2 w_\tau} + \mathcal{H}(\mathcal{P},\delta) \right\}$, $\zeta_\delta \triangleq {\left[\log\left(\frac{\alpha_\delta}{\beta_\delta}\right)\right]^2}$, $$\mathcal{D}_{\tau,j}  \triangleq \frac{1}{w_\tau} \sum_{s=\tau-D-w_\tau + 1}^{\tau-D} \mathbb{E}_{\pi_\tau} \log \left(\frac{\mathcal{P}_{j}(\omega(s))}{\mathcal{P}_{i^*}(\omega(s))}\right),$$ $\mathcal{D}_\tau \triangleq \min_{j \neq i^*} \mathcal{D}_{\tau,j}$, and $\mathcal{H}(\mathcal{P},\delta) = \log M_\delta$ is the \emph{metric entropy}. Further, when $u_t = \mathcal{O}(\sqrt{{t}})$, $v_t = \mathcal{O}(\sqrt{{t}})$, and $\alpha_t = \mathcal{O}(\sqrt{{t}})$, we have $\sum_{\tau=\alpha_t}^{t-1} \Pr\{\mathcal{E}_{\delta,\tau}\} \preceq (t-\alpha_t) S_{t,\delta}$, where $S_{t,\delta} \triangleq \exp\left\{- {\phi_{\tau,t,\delta}} + \mathcal{H}(\mathcal{P},\delta) \right\}$.
In the above, $\phi_{\tau,t,\delta} \triangleq 2 \zeta_\delta \left[\min_{\alpha_t \leq \tau \leq t} \mathcal{D}_\tau\right]^2 N_{[\alpha_t:t]}$, $N_{[\alpha_t:t]} \triangleq \min_{\alpha_t \leq \tau \leq t} w_\tau$.
\end{lemma}
\emph{Proof:} See Appendix \ref{app:prob_error}. $\blacksquare$

From the above lemma, we have that the error goes to zero exponentially fast as $\tau \rightarrow \infty$. The fact that $\sum_{\tau=\alpha_t}^{t-1} \Pr\{\mathcal{E}_{\delta,\tau}\} \preceq (t-\alpha_t) S_{t,\delta} \rightarrow 0$ exponentially fast as $t \rightarrow \infty$ will be used later in the paper to prove the almost sure convergence of the algorithm to the optimal. Now, it remains to find an upper bound on the first and the last term in \eqref{eq:mcdiarmid_pac1}. The following theorem uses the \textbf{Assumption 3}, \eqref{eq:mcdiarmid_pac1} and \eqref{eq:peupperbound_stationary_lemma} to provide a PAC result for the above algorithm in terms of the $\beta_1$ coefficient. 

 \begin{thrm} \label{thm:mainresult1}
Under \textbf{Assumptions 1-3}, for the proposed \textbf{Algorithm} with $\epsilon_0 = (c + 1) \Delta_{\pi,\mathcal{P}_{i^{*}}} + \psi_t(\delta) + \frac{{\alpha_t}{(p_{\text{max},k} -  p_{\text{min},k})}}{t-\alpha_t} + \epsilon$, $\epsilon_k = Q_{\texttt{up}}(t) + \epsilon$, $k=1,2,\ldots, K$, and some finite positive constants $V$, $C$ and $c$, the following holds.
\begin{enumerate}
\item For every $\epsilon > 0$, with a probability of at least $1-\gamma_0$, 
\begin{eqnarray} \label{eq:obj_mainres}
&\frac{1}{t}\sum_{\tau=0}^{t-1} p_0(\tau) \leq&  p^{(opt)} + (c + 1) \Delta_{\pi,\mathcal{P}_{i^{*}}} + \psi_t(\delta) \nonumber \\
&&+ \frac{{\alpha_t}{(p_{\text{max},k} -  p_{\text{min},k})}}{t-\alpha_t} + \epsilon
\end{eqnarray}
provided $t \in \mathcal{T}_{t,0}$. Here, $\gamma_0 > \beta_0^*$, $\beta_0^* \triangleq (t - \alpha_t)\left[ \beta_{\texttt{ADPP},0}(u_t,\alpha_t \left \vert \right. \mathcal{E}_{[\alpha_t:t]}) + S_{t,\delta} \right]$, where $S_{t,\delta}$ is as defined in earlier.

\item For every $\epsilon > 0$, with a probability of at least $1-\gamma_1$, 
\begin{equation} \label{eq:constraint_mainres}
\frac{1}{t}\sum_{\tau=0}^{t-1} p_k(\tau) \leq  c_k + Q_{\texttt{up}}(t) + \frac{{\alpha_t}{(p_{\text{max},k} -  p_{\text{min},k})}}{t-\alpha_t} + \epsilon,
\end{equation}
$k = 1,2,\ldots,K$, \\ provided $t \in \mathcal{T}_{t,1}$. Here $\gamma_1 > \beta_{1}^*$, where $$\beta_1^* \triangleq (t - \alpha_t)\left[ \max_{k\neq 0 }\beta_{\texttt{ADPP},k}(u_t,\alpha_t \left \vert \right. \mathcal{E}_{[\alpha_t:t]}) + S_{t,\delta}\right].$$

\end{enumerate}
In the above, $$\mathcal{T}_{t,i} \triangleq \left\{t: (t - \alpha_t) > \frac{(\Delta p)_{\text{max},0} u_t}{\sqrt{2} \epsilon } \sqrt{\log\left(\frac{u_t}{\gamma_i - \beta_i^*}\right)}\right\},$$ $i \in \{0,1\}$, $\Delta_{\pi,\mathcal{P}_{i^{*}}} = \max_{k=0,1,2,\ldots,K} b_{\text{max},k}  (d_{\pi,\mathcal{P}_{i^{*}}} + \nu)$, and 
\begin{eqnarray}
&\psi_t(\delta) \triangleq& \frac{V(c+1){\bar{J}}_t + \bar{H}_t + C/t}{V}+  \frac{1+2D}{tV}\sum_{\tau=0}^{t-1} B_\tau P_{e,\texttt{up}}^{(\tau)} \nonumber \\
&&+ \frac{p_{\text{max},0}}{t} \sum_{\tau=0}^{t-1} P_{e,\texttt{up}}^{(\tau)} + \frac{\rho}{V t} \sum_{\tau=0}^{t-1} \tau P_{e,\texttt{up}}^{(\tau)},
\end{eqnarray}
where $\rho \triangleq \sum_{k=1}^K (p_{\text{max},k} - c_k)^2$, $\bar{J}_t \triangleq \max_{0\leq k \leq K} p_{\text{max},k} \left(\frac{1}{t}\sum_{\tau = 0}^{t-1}\norm{\pi_\tau - \pi}_1 + \delta \right)$, $\bar{H}_t \triangleq \frac{1+2D}{t}\sum_{\tau = 0}^{t-1} B_\tau$. Further, 
$\mathcal{D}_{\tau,j}$, 
$\mathcal{D}_\tau$, $\zeta_\delta$, and $P_{e,\texttt{up}}^{(\tau)}$ are as defined in Lemma \ref{lemm:prob_error}.
Also, 
$Q_{\texttt{up}}(t)\triangleq \sqrt{\frac{V F}{t} + \frac{\Gamma_t}{ t^2}}$
and $\Gamma_t \triangleq V(c + 1) (\Delta_{\pi,\mathcal{P}_{i^{*}}} + {{\bar{J}_t}) + \bar{H}_t + C} + (1 + 2D) \sum_{\tau=0}^{t-1} B_\tau P_{e,\texttt{up}}^{(\tau)} + p_{\text{max},0} \sum_{\tau=0}^{t-1} P_{e,\texttt{up}}^{(\tau)}+ {\rho} \sum_{\tau=0}^{t-1} \tau P_{e,\texttt{up}}^{(\tau)}$ and $p_{\text{max,k}}$, $k=1,2,\ldots,K$ is as defined earlier.
\end{thrm}
\emph{Proof:} See the Appendix \ref{app:mainresult1}.  $\blacksquare$

The above result can be used to provide \emph{almost sure} convergence as well as finite sample complexity result provided we show that the $\beta_1$ mixing coefficient decays sufficiently fast. This requires us to prove a bound on $\beta_{\texttt{ADPP},k}$. First, we consider a special case of the centralized scheme, i.e., $D=0$. Then, we extend the proof to any $D \geq 0$. The details of this are provided next.

\section{Bound On the Mixing Coefficient} \label{sec:mixing_bound}
By using the Pinsker's inequality that relates the total variational norm and the mutual information, we have the following bound \cite{pinsker}
\begin{equation}
\beta_{\texttt{ADPP},k}(s, \alpha_t \left \vert \right. \mathcal{E}_{[\alpha_t:t]}^c) \leq \sup_{t \geq \alpha_t} \sqrt{\frac{I(X_{k,t};X_{k,t-s} \left \vert \right. \mathcal{E}_{[\alpha_t:t]}^c)}{2}},
\end{equation}
where $X_{k,t} \triangleq p_k(t)$, $I(X_{k,t};X_{k,t-s} \left \vert \right. \mathcal{E}_{[\alpha_t:t]}^c)$ is the mutual information between random variables $p_k(t)$ and $p_k(t-s)$, $k=0,1,2,\ldots,K$ conditioned on $\mathcal{E}_{[\alpha_t:t]}^c$, and any $s \in \mathbb{N}$. Later, we use $s = u_t$, as required. Thus, proving an upper bound on $\beta_{\texttt{ADPP},k}(s, \alpha_t \left \vert \right. \mathcal{E}_{[\alpha_t:t]}^c)$ amounts to finding an upper bound on the conditional mutual information. To present our results, we use the following notations. Let $\mathbf{X}_t \triangleq (X_{0,t},X_{1,t},\ldots,X_{K,t})$, $\mathbf{X}_{\neq k,t} \triangleq (X_{1,t},X_{2,t}\ldots,X_{k-1,t},X_{k+1,t},\ldots,X_{K,t})$, and as before, $\mathbf{Q}_t \triangleq (Q_1(t),Q_2(t),\ldots,Q_K(t))$. We first note that 
\begin{eqnarray}
I(\mathbf X_{t};\mathbf X_{t-s}\left \vert \right. \mathcal{E}_{[\alpha_t:t]}^c) \hspace{-0.3cm} &=& \hspace{-0.2cm} I(X_{k,t};\mathbf X_{t-s}\left \vert \right. \mathcal{E}_{[\alpha_t:t]}^c) + \nonumber \\ && \hspace{-0.2cm} I(X_{\neq k,t};\mathbf X_{t-s} \left \vert \right. X_{k,t}, \mathcal{E}_{[\alpha_t:t]}^c) \nonumber\\
&=& \hspace{-0.2cm} I(X_{k,t};X_{k, t-s}\left \vert \right. \mathcal{E}_{[\alpha_t:t]}^c) +\nonumber \\ && \hspace{-0.2cm} I(X_{k,t};\mathbf X_{\neq k, t-s} \left \vert \right. X_{k, t-s}, \mathcal{E}_{[\alpha_t:t]}^c) +\nonumber \\ && \hspace{-0.2cm} I(\mathbf X_{\neq k,t};\mathbf X_{t-s} \left \vert \right. X_{k,t},\mathcal{E}_{[\alpha_t:t]}^c) \nonumber \\
&\geq& \hspace{-0.2cm} I(X_{k,t};X_{k, t-s}\left \vert \right. \mathcal{E}_{[\alpha_t:t]}^c),
\end{eqnarray}
where the last inequality follows from the fact that the mutual information is non-negative. Thus, we have
\begin{equation} \label{eq:betaone_bound_premitive}
\beta_{\texttt{ADPP},k}(s, \alpha_t \left \vert \right. \mathcal{E}_{[\alpha_t:t]}^c) \leq \sup_{t \geq \alpha_t} \sqrt{\frac{I(\mathbf X_{t};\mathbf X_{t-s}\left \vert \right. \mathcal{E}_{[\alpha_t:t]}^c)}{2}}.
\end{equation}
Let $\mathcal{Q}_t$ be the set of all vectors that $\mathbf{Q}_t$ takes at time $t$. Also, let $\mathcal{M}_t : \mathcal{Q}_t \rightarrow \{1,2,\ldots,F\}$ be the rule induced by the \texttt{ADPP} algorithm that determines the strategy given the queue at time $t$. In order to obtain an upper bound on the mutual information, we state the following assumption about the conditional distribution of the process $\omega(t)$. 

\textbf{Assumption 4:} For some $\kappa > 0$, $\mathbf{Q}_t \in \mathcal{Q}_t$, and for all $t \in \mathbb{N}$, we assume that the following bound is satisfied
\begin{equation}
\sup_{x,m,m^{'}}\frac{\Pr\{\mathbf X_t = x \left \vert \right. \mathcal{M}_t(\mathbf{Q}_t) = m, \mathcal{E}_{[\alpha_t:t]}^c\}}{\Pr\{\mathbf  X_t = x \left \vert \right. \mathcal{M}_t(\mathbf{Q}_t) = m^{'}, \mathcal{E}_{[\alpha_t:t]}^c\}} \leq e^{\kappa}.
\end{equation}
 Note that a lower value of $\kappa$ signifies the fact that the channel is noisy. For example, when $\kappa = 0$, we have uniform conditional distribution for all $m$ and $x$ leading to a completely noisy channel from $Q_\tau$ to $X_\tau$. 
Next, we present an upper bound on $\beta_{\texttt{ADPP},k}(s, \alpha_t \left \vert \right. \mathcal{E}_{[\alpha_t:t]}^c)$ for the $D=0$ case (centralized scheme).

\subsection{Bound on $\beta_{\texttt{ADPP},k}(s, \alpha_t \left \vert \right. \mathcal{E}_{[\alpha_t:t]}^c)$ when $D=0$}
\begin{figure}[h]
    \centering
    \includegraphics[width=8.0cm, height = 3.2cm]{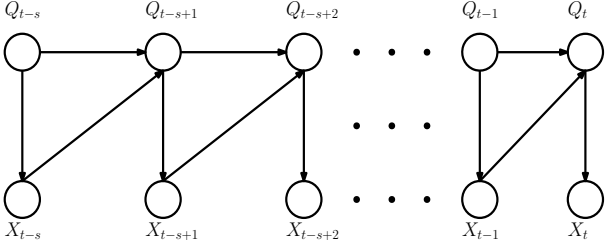}
    \caption{Figure shows the graphical model corresponding to the \texttt{ADPP} algorithm with $D=0$ and time slots from $t-s \geq \alpha_t$ to $t$.}
    \label{fig:graphicalmodel_1}
\end{figure}

In order to get insights on the proof of bounding the $\beta_1$ coefficient for the general scenario of $D \geq 0$, we first consider the centralized scheme, i.e., $D=0$, and later we provide proofs and results for the $D>0$ case. For $D=0$, the queue update in the vector form becomes 
\begin{equation}
\mathbf{Q}_{t+1} = \max\left\{\mathbf{Q}_{t} + \mathbf X_{\neq 0, t} - \mathbf C, 0\right\}
\end{equation}
where $\mathbf  C \triangleq (c_1,c_2,\ldots,c_K)$. Recall that \textbf{Step 2} of the \textbf{Algorithm} uses $\mathbf{Q}_{t}$ and the output from \textbf{Step 1} to find a pure strategy in a deterministic fashion that maximizes an upper bound on the drift-plus-penalty expression. Thus, the strategy is a deterministic function of the queue. Note that conditioned on the event $\mathcal{E}_{[\alpha_t:t]}^c$, the output of \textbf{Step 1} is $i^*$ for all time slots $\tau \in \{\alpha_t, \ldots, t\}$. Conditioned on $\mathcal{E}_{[\alpha_t:t]}^c$, this leads to the following Markov chain model
\begin{equation}
(\mathbf{Q}_{\alpha_t}, \mathbf X_{\alpha_t}) \longrightarrow (\mathbf{Q}_{\alpha_t + 1}, \mathbf X_{\alpha_t + 1})  \longrightarrow  \ldots \longrightarrow  (\mathbf{Q}_{t}, \mathbf X_{t}). \nonumber
\end{equation}
Fig. \ref{fig:graphicalmodel_1} depicts the graphical model representation of the above. In order to prove an upper bound on the mutual information, we use the Strong Data Processing Inequality (SDPI) for the graphical model shown in Fig. \ref{fig:graphicalmodel_1}. 
  
Note that the \textbf{Assumption 4} facilitates the analysis of the $\beta_1$ mixing coefficient, and is also related to the differential privacy constraint in \cite{duchi2013local}. The following theorem provides an upper bound on the mixing coefficient. 
\begin{thrm} \label{thm:d1_betabound}
Given \textbf{Assumption 4}, for $D=0$, $\kappa < \log 3$, and for any $t \geq s \geq \alpha_t$, an upper bound on the $\beta_1$ mixing coefficient is given by 
\begin{equation}
\beta_{\texttt{ADPP},k}(s, \alpha_t \left \vert \right. \mathcal{E}_{[\alpha_t:t]}^c)  \leq {\frac{\theta^{(s-1)/2}}{\sqrt{2}} \left[\log \mu\right]}, k = 0,1,2,\ldots,K
\end{equation}
where $\mu \triangleq F \abs{\Omega} (K+1)$ is the number of possible values that $\mathbf  X_t$ can take, $t \in \mathbb{N}$, and $\theta \triangleq \max\left\{\frac{(e^{\kappa}-1)}{2}, \frac{1}{2}\right\} < 1$.  
\end{thrm}
\emph{Proof:} See the Appendix \ref{app:d1_betabound}.  $\blacksquare$

Note that $s = u_t$, and suppose $u_t$ grows with $t$, then the Theorem says that the mixing coefficient goes down to zero exponentially fast with $t$. Thus, we have the following important corollary. 
\begin{corol} \label{corr:mixing_d1}
Given \textbf{Assumption 4}, for $D=0$, $u_t = \mathcal{O}(\sqrt{t})$, $\kappa < \log 3$, and for any $t \geq s \geq \alpha_t$, an upper bound on the $\beta_1$ mixing coefficient is given by 
\begin{equation} \label{eq:mixing_d1}
\beta_{\texttt{ADPP},k}(u_t, \alpha_t \left \vert \right. \mathcal{E}_{[\alpha_t:t]}^c)  \preceq {\frac{(\theta^{\mathcal{O}(\sqrt{t})/2})}{\sqrt{2}} \left[\log \mu\right]},
\end{equation}
where $\mu \triangleq F \abs{\Omega} (K+1)$ is the number of possible values that $\mathbf X_t$ can take, $t \in \mathbb{N}$, and $\theta \triangleq \max\left\{\frac{(e^{\kappa}-1)}{2}, \frac{1}{2}\right\} < 1$.  
\end{corol}

A finite time bound can easily be obtained by substituting the upper bound of Theorem \ref{thm:d1_betabound} in Theorem \ref{thm:mainresult1}. However, in order to get more insights into the main result of this paper, we will look at the asymptotic in the following subsection.

\subsubsection{Asymptotics}
Note that when $D=0$, the authors in \cite{wei2015sample} prove the convergence of the algorithm to the optimal in probability. Here, we use a different approach compared to \cite{wei2015sample} to show an \emph{almost sure} as well as a high probability convergence of the  the proposed \texttt{ADPP} algorithm to the optimal, when $D=0$. By using the insights obtained here, we generalize the result to an arbitrary $D \geq 0$ in the subsequent subsections. First, in the following lemma, we provide a high probability guarantees of the \texttt{ADPP} algorithm when $D=0$ and $t \rightarrow \infty$. 
\begin{lemma} \label{lem:main1}
Under \textbf{Assumptions 1-4}, for the proposed \textbf{Algorithm} with $D=0$, $\alpha_t = \mathcal{O}(\sqrt{t})$, $w_t = \mathcal{O}(\sqrt{t})$, $V=\mathcal{O}(\sqrt{t})$, $\kappa < \log 3$, and some finite positive constant $c$, the following holds.
\begin{itemize}
\item For every $\epsilon > 0$, we have  
\begin{eqnarray} \label{eq:obj_mainres_hp}
&& \hspace{-2cm}\lim_{t \rightarrow \infty} \Pr \left\{\frac{1}{t}\sum_{\tau=0}^{t-1} p_0(\tau) \leq  p^{(opt)} + (c + 1) \Delta_{\pi,\mathcal{P}_{i^{*}}} \right.  \nonumber \\
&&  \left.+ {(c+1){\bar{J}} } +  \epsilon  \right\} = 1
\end{eqnarray}
\text{and  } $\lim_{t \rightarrow \infty} \Pr \left\{\frac{1}{t}\sum_{\tau=0}^{t-1} p_k(\tau) \leq  c_k  + \epsilon\right\} = 1$, 
k = 1,2,\ldots,K. 
\end{itemize}
In the above, $\Delta_{\pi,\mathcal{P}_{i^{*}}} = \max_{k=0,1,2,\ldots,K} b_{\text{max},k}  (d_{\pi,\mathcal{P}_{i^{*}}} + \nu)$, $$\bar{J} \triangleq \max_{0\leq k \leq K} p_{\text{max},k} \left(\lim_{t \rightarrow \infty}\frac{1}{t}\sum_{\tau = 0}^{t-1}\norm{\pi_\tau - \pi}_1 + \delta \right),$$ 
and $p_{\text{max,k}}$, $k=1,2,\ldots,K$ is as defined earlier.

\end{lemma}
\emph{Proof:} See Appendix \ref{app:as_D1_proof}. $\blacksquare$

The interpretations of the above result will be provided later. Next, we use Lemma \ref{lemm:prob_error} along with the Borel-Cantelli Lemma to provide an almost sure convergence of the \texttt{ADPP} algorithm. 
\begin{thrm} \label{thm:as_1}
Under \textbf{Assumptions 1-4}, for the proposed \textbf{Algorithm} with $D=0$, $\alpha_t = \mathcal{O}(\sqrt{t})$, $w_t = \mathcal{O}(\sqrt{t})$, $V= \mathcal{O}(\sqrt{t})$, $\kappa < \log 3$, and some finite positive constant $c$, the following holds.
\begin{itemize}
\item For every $\epsilon > 0$, \emph{almost surely}, we have  
\begin{equation} \label{eq:obj_mainres_as_D1}
\lim_{t \rightarrow \infty} \frac{1}{t}\sum_{\tau=0}^{t-1} p_0(\tau) \leq  p^{(opt)} + (c + 1) \Delta_{\pi,\mathcal{P}_{i^{*}}} + {(c+1){\bar{J}} } +  \epsilon 
\end{equation}
\begin{equation} \label{eq:constraint_mainres_as_D1}
\text{and  }\lim_{t \rightarrow \infty} \frac{1}{t}\sum_{\tau=0}^{t-1} p_k(\tau) \leq  c_k  + \epsilon, ~k = 1,2,\ldots,K.
\end{equation}
\end{itemize}
In the above, $\Delta_{\pi,\mathcal{P}_{i^{*}}}$, $\bar{J}$, 
and $p_{\text{max,k}}$, $k=1,2,\ldots,K$ are as defined earlier.
\end{thrm}
\emph{Proof:} See Appendix \ref{app:almost_sure_d1}. $\blacksquare$

From Theorems \ref{thm:mainresult1} and \ref{thm:as_1}, it is easy to see that the error can be reduced by reducing $\Delta_{\pi,\mathcal{P}_{i^{*}}}$, which amounts to reducing $d_{\pi,\mathcal{P}_{i^{*}}}$ and $\nu$. Note that $d_{\pi,\mathcal{P}_{i^{*}}} < \delta$ can be reduced by reducing the error in the covering of the probability space $\mathcal{P}_c$. This comes at a cost of increased metric entropy since $\delta$ needs to be reduced. However, as $t \rightarrow \infty$, increased metric entropy does not effect the overall result. Further, a lower value of $\bar J$ signifies lesser error. This is possible only when the rate at which the probability measure $\pi_t$ converges to $\pi$ is ``sufficiently" high. In particular, this is true when $\sum_{\tau = 0}^{t-1} \norm{\pi_\tau - \pi}_{1} = \mathcal{O}(t^\zeta)$, where $\zeta <1$.  In the next subsection, we provide an almost sure as well as high probability result for any $D \geq 0$. 

\subsection{Bound on $\beta_{\texttt{ADPP},k}(s, \alpha_t \left \vert \right. \mathcal{E}_{[\alpha_t:t]}^c)$ when $D \geq 0$}

\begin{figure}[h]
	\centering
	\includegraphics[width=8.5cm, height = 5cm]{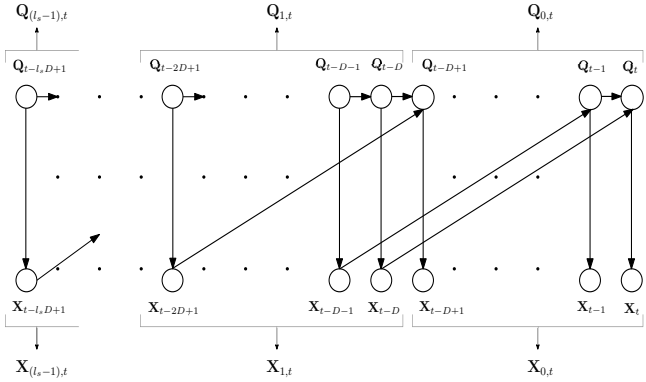}
	\caption{Figure shows the graphical model corresponding to the \texttt{ADPP} algorithm with $D \geq 0$ and time slots from $t-s \geq \alpha_t$ to $t$.}
	\label{fig:graphicalmodel_2}
\end{figure}

As in the previous subsection, we use $s=u_t$. For $D \geq 0$, the queue update in the vector form is given by 
\begin{equation}
\mathbf{Q}_{t+1} = \max\left\{\mathbf{Q}_{t} + \mathbf X_{\neq 0, t - D} - \mathbf C, 0\right\}
\end{equation}
where $\mathbf C \triangleq (c_1,c_2,\ldots,c_K)$ and $\mathbf X_{\neq 0, t - D}$ is as defined earlier in this section. As in the $D=0$ case, we condition on the event $\mathcal{E}_{[\alpha_t:t]}^c$, and therefore, the output of \textbf{Step 1} is $i^*$ for all time slots $\tau \in \{\alpha_t, \ldots, t\}$.\footnote{Recall that $i^*$ is the index corresponding to $\pi_{i^*}$, which is the distribution ``close" to $\pi$.} Define the following shorthand notations $\mathbf Q_{1:n} \triangleq \{\mathbf Q_1, \mathbf Q_2, \ldots, \mathbf Q_n\}$, and $\mathbf X_{1:m} \triangleq \{\mathbf X_1,\mathbf X_2,\ldots \mathbf X_m\}$. Unlike the  $D=0$ case, conditioning on $\mathcal{E}_{[\alpha_t:t]}^c$ leads to the following Markov chain model
\begin{eqnarray}
&&\hspace{-0.7cm}(\mathbf{Q}_{t-l_sD + 1: t-(l_s-1)D}, X_{t-l_sD + 1: t-(l_s-1)D})  \rightarrow  \ldots \nonumber \\
&&\hspace{-0.7cm}\rightarrow (\mathbf{Q}_{t-2D + 1: t-D}, X_{t-2D+1:t-D}) \rightarrow  (\mathbf{Q}_{t-D + 1: t}, X_{t-D+1:t}), \nonumber
\end{eqnarray}
where $l_s \triangleq \lceil \frac{s+1}{D} \rceil$.
Fig. \ref{fig:graphicalmodel_2} depicts the graphical model representation of the above. 
Note that the $i^{th}$ pair in the Markov chain is $(\mathbf{Q}_{t-(i+1)D + 1: t-iD}, X_{t-(i+1)D+1:t-iD})$, $i=0,1,\ldots,l_s-1$. In order to prove an upper bound on the mutual information, we use the Strong Data Processing Inequality (SDPI) for the graphical model shown in Fig. \ref{fig:graphicalmodel_2}. Using the above mentioned Markov property, we need to bound the term $I(\mathbf X_{t};\mathbf X_{t-s}\left \vert \right. \mathcal{E}_{[\alpha_t:t]}^c)$ from \eqref{eq:betaone_bound_premitive}. In the following theorem, we present a bound on the $\beta_1$ mixing coefficient for $D \geq 0$ case.

\begin{thrm} \label{thm:dgeq1_betamixing_mainres}
Given \textbf{Assumption 4}, for $D \geq 0$, $D < \frac{\log 3}{\kappa}$, and for any $t \geq s \geq \max\{\alpha_t, 2D + 1\}$, an upper bound on the $\beta_1$ mixing coefficient is given by 
\begin{equation}
\beta_{\texttt{ADPP},k}(s, \alpha_t \left \vert \right. \mathcal{E}_{[\alpha_t:t]}^c)  \leq {\frac{\theta^{(s-D + 1)/2D}}{\sqrt{2}} \left[\log \mu_D\right]}, 
\end{equation}
$k = 0,1,2,\ldots,K$, where $\mu_D \triangleq D F \abs{\Omega} (K+1)$ is the number of possible values that $X_t$ can take, $t \in \mathbb{N}$, and $\theta \triangleq \max\left\{\frac{(e^{{\kappa}D}-1)}{2}, \frac{1}{2}\right\} < 1$.  
\end{thrm}
\emph{Proof:} For the ease of notation, let $\mathbf{X}_{i,t} \triangleq \mathbf X_{t-(i+1)D+1:t-iD}$ and $\mathbf{Q}_{i,t} \triangleq \mathbf{Q}_{t-(i+1)D+1:t-iD}$. First, in the following, we prove that $I(\mathbf X_{t};\mathbf X_{t-s}\left \vert \right. \mathcal{E}_{[\alpha_t:t]}^c) \leq I(\mathbf{X}_{0,t};\mathbf{X}_{l_s - 1,t}\left \vert \right. \mathcal{E}_{[\alpha_t:t]}^c)$. Consider
\begin{eqnarray}
I(\mathbf X_{t};\mathbf X_{t-s}\left \vert \right. \mathcal{E}_{[\alpha_t:t]}^c) &\stackrel{(a)}{\leq}& I(\mathbf{X}_{0,t};X_{t-s}\left \vert \right. \mathcal{E}_{[\alpha_t:t]}^c) \nonumber\\
&\stackrel{(b)}{\leq}& I(\mathbf{X}_{0,t};\mathbf{X}_{l_s - 1,t}\left \vert \right. \mathcal{E}_{[\alpha_t:t]}^c),
\end{eqnarray}
where $(a)$ and $(b)$ follow from the definitions of $\mathbf X_t \in \mathbf{X}_{0,t}$ and $\mathbf X_{t-s} \in \mathbf{X}_{l_s-1,t}$, and the fact that the mutual information is non-negative. We need to upper bound $ I(\mathbf{X}_{0,t};\mathbf{X}_{l_s - 1,t}\left \vert \right. \mathcal{E}_{[\alpha_t:t]}^c)$, which is obtained in a manner similar to the $D=0$ case, as explained next. Since $\mathbf{X}_{l_s - 1,t}  \rightarrow  {\mathbf{Q}_{0,t}} \rightarrow \mathbf{X}_{0,t} $ forms a Markov chain, we obtain the following bound from the SDPI 
\begin{equation} \label{eq:sdpi_dgeq1_bound1}
I(\mathbf{X}_{l_s - 1,t};\mathbf{X}_{0,t}\left \vert \right. \mathcal{E}_{[\alpha_t:t]}^c) \leq \eta_{\texttt{ch}_1} I(\mathbf{X}_{l_s - 1,t};{\mathbf{Q}_{0,t}}\left \vert \right. \mathcal{E}_{[\alpha_t:t]}^c),
\end{equation}
where $\eta_{\texttt{ch}_1}$ is the Dobrushin's contraction coefficient for the channel from ${\mathbf{Q}_{0,t}}$ to ${\mathbf{X}_{0,t}}$ defined as  
\begin{eqnarray} \label{eq:defn_etach1}
&&\hspace{-1cm}\eta_{\texttt{ch}_1} \triangleq \sup_{\gamma \neq \gamma^{'}} \norm{\Pr\left\{{\mathbf{X}_{0,t}} \left \vert \right.{\mathbf{Q}_{0,t}} = \gamma, {\mathcal{E}_{[\alpha_t:t]}^c} \right\} \nonumber \\
&&\hspace{1cm}- \Pr\left\{{\mathbf{X}_{0,t}} \left \vert \right. {\mathbf{Q}_{0,t}} = \gamma^{'}, {\mathcal{E}_{[\alpha_t:t]}^c} \right\}}_{\texttt{TV}}.
\end{eqnarray}
In the above, $\gamma$ and $\gamma^{'}$ represent the vector values taken by ${\mathbf{Q}_{0,t}}$. It will {be} shown later that $\eta_{\texttt{ch}_1} < 1$. Note that by simple data processing inequality, we have from \eqref{eq:sdpi_dgeq1_bound1} that 
\begin{eqnarray}
I(\mathbf{X}_{l_s - 1,t};\mathbf{X}_{0,t}\left \vert \right. \mathcal{E}_{[\alpha_t:t]}^c) \hspace{-0.2cm} &\leq& \hspace{-0.2cm} \eta_{\texttt{ch}_1} I(\mathbf{X}_{l_s - 1,t};{\mathbf{Q}_{0,t}}\left \vert \right. \mathcal{E}_{[\alpha_t:t]}^c)\nonumber \\ &&\hspace{-1.8cm} \leq  \eta_{\texttt{ch}_1} I(\mathbf{X}_{l_s - 1,t};{\mathbf{Q}_{1,t}}, {\mathbf{X}_{1,t}}\left \vert \right. \mathcal{E}_{[\alpha_t:t]}^c).
\end{eqnarray}
The first inequality above follows from the fact that ${\mathbf{Q}_{0,t}}$ is a deterministic function of ${\mathbf{X}_{1,t}}$ and ${\mathbf{Q}_{1,t}}$. Since $\mathbf{X}_{l_s - 1,t}  \rightarrow  ({\mathbf{Q}_{2,t}},{\mathbf{X}_{2,t}}) \rightarrow ({\mathbf{Q}_{1,t}},{\mathbf{X}_{1,t}})$ forms a Markov chain, the above can be further bounded as follows (see Fig. \ref{fig:graphicalmodel_2})
\begin{eqnarray} \label{eq:mutbound_dobrushin_2}
&&\hspace{-0.7cm}I(\mathbf{X}_{l_s - 1,t};\mathbf{X}_{0,t}\left \vert \right. \mathcal{E}_{[\alpha_t:t]}^c) \leq \eta_{\texttt{ch}_1} I({\mathbf{X}_{l_s - 1,t}};{\mathbf{Q}_{1,t}}, {\mathbf{X}_{1,t}}\left \vert \right. \mathcal{E}_{[\alpha_t:t]}^c) \nonumber \\
&&\hspace{1.5cm}\leq \eta_{\texttt{ch}_1} \eta_{\texttt{ch}_2} I({\mathbf{X}_{l_s - 1,t}};{\mathbf{Q}_{2,t}}, {\mathbf{X}_{2,t}}\left \vert \right. \mathcal{E}_{[\alpha_t:t]}^c),
\end{eqnarray}
where $\eta_{\texttt{ch}_2}$ is the Dobrushin's coefficient for the channel $({\mathbf{Q}_{2,t}}, {\mathbf{X}_{2,t}})$ to $({\mathbf{Q}_{1,t}}, {\mathbf{X}_{1,t}})$ defined as 
\begin{eqnarray}
\eta_{\texttt{ch}_2} \hspace{-0.3cm} &\triangleq& \hspace{-0.6cm}\sup_{(p,q) \neq (p^{'}, q^{'})} \left \vert \left \vert \Pr\left\{{\mathbf{X}_{1,t}}, {\mathbf{Q}_{1,t}} \left \vert \right.{\mathbf{X}_{2,t}} = p, {\mathbf{Q}_{2,t}} = q, {\mathcal{E}_{[\alpha_t:t]}^c} \right\} \right.\right.\nonumber \\ 
&& \hspace{-0.8cm} \left. \left.  - \Pr\left\{{\mathbf{X}_{1,t}}, {\mathbf{Q}_{1,t}} \left \vert \right.{\mathbf{X}_{2,t}} = p^{'}, {\mathbf{Q}_{2,t}} = q^{'}, {\mathcal{E}_{[\alpha_t:t]}^c} \right\}\right \vert \right \vert_{\texttt{TV}}.
\end{eqnarray}
Note that $\mathbf{X}_{l_s - 1,t}  \rightarrow  ({\mathbf{Q}_{j,t}},{\mathbf{X}_{j,t}}) \rightarrow ({\mathbf{Q}_{j-1,t}},{\mathbf{X}_{j-1,t}})$ forms a Markov chain for all $j=2,3,\ldots,l_s - 2$. The corresponding Dobrushin coefficient is given by 
\begin{eqnarray}
&&\hspace{-0.7cm}\eta_{\texttt{ch}_j} \triangleq \nonumber \\ 
&&\hspace{-0.7cm}\sup_{(p,q) \neq (p^{'}, q^{'})} \left \vert \left \vert \Pr\left\{{\mathbf{X}_{j-1,t}}, {\mathbf{Q}_{j-1,t}} \left \vert \right.{\mathbf{X}_{j,t}} = p, {\mathbf{Q}_{j,t}} = q, {\mathcal{E}_{[\alpha_t:t]}^c} \right\} \right.\right.\nonumber \\ && \left. \left.  - \Pr\left\{{\mathbf{X}_{j-1,t}}, {\mathbf{Q}_{j-1,t}} \left \vert \right.{\mathbf{X}_{j,t}} = p^{'}, {\mathbf{Q}_{j,t}} = q^{'}, {\mathcal{E}_{[\alpha_t:t]}^c} \right\}\right \vert \right \vert_{\texttt{TV}} \nonumber \\
&&\hspace{-0.7cm}=\sup_{(p,q) \neq (p^{'}, q^{'})} \left \vert \left \vert \left( \Pr\left\{{\mathbf{Q}_{j-1,t}} \left \vert \right.{\mathbf{X}_{j,t}} = p, {\mathbf{Q}_{j,t}} = q, {\mathcal{E}_{[\alpha_t:t]}^c} \right\} \right. \right. \right. \nonumber \\ 
&& \hspace{1.5cm} \times \left. \Pr\left\{{\mathbf{X}_{j-1,t}} \left \vert \right. {\mathbf{Q}_{j-1,t}} = q,{\mathcal{E}_{[\alpha_t:t]}^c} \right\} \right) \nonumber \\ 
&& \hspace{0.7cm} - \left(\Pr\left\{{\mathbf{Q}_{j-1,t}} \left \vert \right.{\mathbf{X}_{j,t}} = p^{'}, {\mathbf{Q}_{j,t}} = q^{'}, {\mathcal{E}_{[\alpha_t:t]}^c} \right\}\right. \nonumber \\
&&\hspace{1.5cm}\left. \left. \left.\times \Pr\left\{{\mathbf{X}_{j-1,t}} \left \vert \right. {\mathbf{Q}_{j-1,t}} = q^{'}, {\mathcal{E}_{[\alpha_t:t]}^c} \right\} \right) \right \vert \right \vert_{\texttt{TV}}. \label{eq:eta_chj_equality}
\end{eqnarray}
Using these in \eqref{eq:mutbound_dobrushin_2}, and applying the bound repeatedly, we get  
\begin{eqnarray}
&&\hspace{-0.7cm}I(\mathbf{X}_{l_s - 1,t};\mathbf{X}_{0,t}\left \vert \right. \mathcal{E}_{[\alpha_t:t]}^c) \nonumber \\
&&\hspace{-0.4cm}\leq \eta_{\texttt{ch}_1} \eta_{\texttt{ch}_2} I(\mathbf{X}_{l_s - 1,t};{\mathbf{Q}_{2,t}}, {\mathbf{X}_{2,t}}\left \vert \right. \mathcal{E}_{[\alpha_t:t]}^c) \nonumber \\
&&\hspace{-0.4cm}\leq \left[\prod_{j=1}^{l_s-2} \eta_{\texttt{ch}_j} \right] I(\mathbf{X}_{l_s - 1,t};{\mathbf{Q}_{l_s - 2,t}}, {\mathbf{X}_{l_s - 2,t}}\left \vert \right. \mathcal{E}_{[\alpha_t:t]}^c).
\end{eqnarray}
But, $I(\mathbf{X}_{l_s - 1,t};{\mathbf{Q}_{l_s - 2,t}},{\mathbf{X}_{l_s - 2,t}}\left \vert \right. \mathcal{E}_{[\alpha_t:t]}^c) \leq H(\mathbf{X}_{l_s - 2,t}) \leq \log N_D$, where $N_D \triangleq D F \abs{\Omega} (K+1)$ is the number of possible values that $\mathbf{X}_{l_s - 2,t}$ can take. Using this in the above, we get
\begin{equation} \label{eq:dgeq1_mutualinfo_bound_incomplete}
I(\mathbf{X}_{0,t};\mathbf{X}_{l_s - 1,t}\left \vert \right. \mathcal{E}_{[\alpha_t:t]}^c) \leq \left[\prod_{j=1}^{l_s-2} \eta_{\texttt{ch}_j} \right] \log N_D.
\end{equation}
Next, in Lemmas \ref{lemm:etach1_bound} and \ref{lemm:etachj_bound}, we prove an upper bound on $\eta_{\texttt{ch}_j}$ for every $j = 1,2,\ldots, l_s - 2$. 
\begin{lemma} \label{lemm:etach1_bound}
Under \textbf{Assumption $4$}, for every $D \geq 0$, and $D < \frac{\log 3}{\kappa}$, we have the  following upper bound on $\eta_{\texttt{ch}_1}$ 
\begin{equation}
\eta_{\texttt{ch}_1} \leq \max\left\{\frac{(e^{{\kappa}D}-1)}{2}, \frac{1}{2}\right\}  < 1.
\end{equation}
\end{lemma}
\emph{Proof:} See Appendix \ref{app:etach1_bound}. $\blacksquare$

\begin{lemma} \label{lemm:etachj_bound}
Under \textbf{Assumption $4$}, for every $D \geq 0$, and $D < \frac{\log 3}{\kappa}$, we have the  following upper bound  
\begin{equation}
\eta_{\texttt{ch}_j} \leq \max\left\{\frac{(e^{{\kappa}D}-1)}{2}, \frac{1}{2}\right\}  < 1, ~j =1,2,\ldots,l_s-2.
\end{equation}
\end{lemma}
\emph{Proof:} See Appendix \ref{app:etachj_bound}. $\blacksquare$ \\
Since there are $l_s-1$ terms in the overall product, we need $l_s - 1\geq 1 \Rightarrow s \geq 2D+1$. Using the bounds in Lemmas \ref{lemm:etach1_bound} and \ref{lemm:etachj_bound} in \eqref{eq:dgeq1_mutualinfo_bound_incomplete}, and substituting the result in \eqref{eq:betaone_bound_premitive}, we get the desired result, which completes the proof of Theorem \ref{thm:dgeq1_betamixing_mainres}. $\blacksquare$

The above result says that for a given $\kappa$, the \texttt{ADPP} algorithm converges to the optimal provided the delay $D$ in the available samples at each node is bounded by a constant $\frac{\log 3}{\kappa}$.

\subsubsection{Almost sure convergence}
Using Lemmas \ref{lemm:etach1_bound} and \ref{lemm:etachj_bound}, and the main result in Theorem \ref{thm:dgeq1_betamixing_mainres}, the following result can be obtained in a fashion similar to the $D=0$ case. In the following lemma, we provide a high probability guarantees of the \texttt{ADPP} algorithm for a general $D \geq1$ and as $t \rightarrow \infty$. 
\begin{lemma} \label{lem:main2}
Under \textbf{Assumptions 1-3}, for the proposed \textbf{Algorithm} with ${D \geq 0}$, $\alpha_t = \mathcal{O}(\sqrt{t})$, $w_t = \mathcal{O}(\sqrt{t})$, $\kappa < \frac{\log 3}{D}$, and some finite positive constant $c$, the following holds.
\begin{itemize}
\item For every $\epsilon > 0$, we have  
\begin{eqnarray} \label{eq:obj_mainres_as_Dgeq1}
&&\hspace{-1cm}\lim_{t \rightarrow \infty} \Pr \left\{\frac{1}{t}\sum_{\tau=0}^{t-1} p_0(\tau) \leq  p^{(opt)} + (c + 1) \Delta_{\pi,\mathcal{P}_{i^{*}}} \right.\nonumber \\
&&\hspace{0.5cm}+\left. {(c+1){\bar{J}} } +  \epsilon\right\} = 1
\end{eqnarray}
\begin{eqnarray} \label{eq:constraint_mainres_Dgeq1}
&&\hspace{-1cm}\text{and  }\lim_{t \rightarrow \infty} \Pr \left\{\frac{1}{t}\sum_{\tau=0}^{t-1} p_k(\tau) \leq  c_k  + \epsilon\right\} = 1,  \\
&&\hspace{0.5cm}k = 1,2,\ldots,K. \nonumber
\end{eqnarray}
\end{itemize}
In the above, $\Delta_{\pi,\mathcal{P}_{i^{*}}} = \max_{k=0,1,2,\ldots,K} b_{\text{max},k}  (d_{\pi,\mathcal{P}_{i^{*}}} + \nu)$, $$\bar{J} \triangleq \max_{0\leq k \leq K} p_{\text{max},k} \left(\lim_{t \rightarrow \infty}\frac{1}{t}\sum_{\tau = 0}^{t-1}\norm{\pi_\tau - \pi}_1 + \delta \right),$$ 
and $p_{\text{max,k}}$, $k=1,2,\ldots,K$ is as defined earlier.

\end{lemma}
\emph{Proof:} The proof is similar to the proof of Lemma \ref{lem:main1}, and hence omitted. $\blacksquare$

Note that the effect of $D$ is captured through the requirement of $\kappa$, i.e., $\kappa < \frac{\log 3}{D}$. In particular, large values of the delay $D$ require stringent constraint on the ``noisyness" of the channel, i.e., a lower value of $\kappa$. However, in many cases, $\kappa$ need not be low, and hence the $\beta_1$ coefficient may not converge to zero exponentially. Thus, the \texttt{ADPP} algorithm may not be asymptotically optimal. In the next subsection, we consider the general case of $D>1$. We show that the \texttt{ADPP} algorithm converges to the optimal in the almost sure sense. 
\begin{thrm} \label{thm:as_2}
Under \textbf{Assumptions 1-4}, for the proposed \textbf{Algorithm} with $D {\geq} 1$, $\alpha_t = \mathcal{O}(\sqrt{t})$, $w_t = \mathcal{O}(\sqrt{t})$, $\kappa \leq \frac{\log 3}{D}$, and some finite positive constant $c$, the following holds.
\begin{itemize}
\item For every $\epsilon > 0$, \emph{almost surely}, we have  
\begin{equation} \label{eq:obj_mainres_Dgeq1}
\lim_{t \rightarrow \infty} \frac{1}{t}\sum_{\tau=0}^{t-1} p_0(\tau) \leq  p^{(opt)} + (c + 1) \Delta_{\pi,\mathcal{P}_{i^{*}}} + {(c+1){\bar{J}} } +  \epsilon 
\end{equation}
\begin{equation} \label{eq:constraint_mainres_dgeq1}
\text{and  }\lim_{t \rightarrow \infty} \frac{1}{t}\sum_{\tau=0}^{t-1} p_k(\tau) \leq  c_k  + \epsilon, ~k = 1,2,\ldots,K.
\end{equation}
\end{itemize}
In the above, $\Delta_{\pi,\mathcal{P}_{i^{*}}}$, $\bar{J}$, 
and $p_{\text{max,k}}$, $k=1,2,\ldots,K$ are as defined earlier.
\end{thrm}
\emph{Proof:} The proof is similar to the proof of Theorem \ref{thm:as_1}, and hence omitted. $\blacksquare$

In the next section, we present the simulation results to corroborate some of the observations made in the paper. 

\begin{figure}[h]
    \centering
    \includegraphics[width=7.0cm, height = 4.5cm]{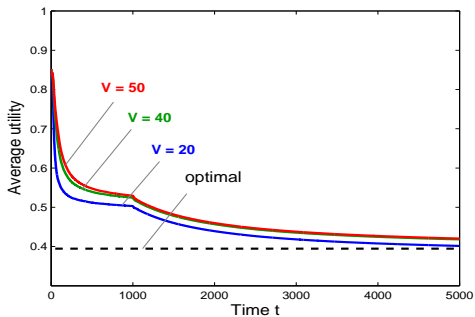}
    \caption{Figure shows the plot of the average utility versus time.}
    \label{fig:utility_vs_time}
\end{figure}
\begin{figure}[h]
    \centering
    \includegraphics[width=7.0cm,height = 4.5cm]{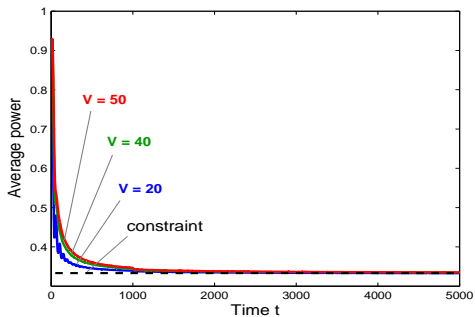}
    \caption{Figure shows the plot of the average power versus time.}
    \label{fig:power_vs_time}
\end{figure}

\section{Simulation Results} \label{sec:sims}
For the simulation setup, we consider the $3$ sensors example of Sec. \ref{sec:sys_model}. The problem is to maximize the average of the utility in \eqref{eq:example} 
subject to an average power constraint of $1/3$. {Here, the utility is the negative of the cost.} The probability measure $\pi_t$ is chosen from a set of $8$ distributions, and converges to $\{0.1,0.7,0.1,0.1\}$. 
Due to lack of space, we skip the details of the distribution that is used in the transient time. The optimal value of this is $p^{(opt)} = 0.394$. When $\alpha_i(t) = 1$, $i=1,2,3$, a power of $1$ watt each is consumed. Figures \ref{fig:utility_vs_time} and \ref{fig:power_vs_time} show the plots of utility and penalty averaged over $1000$ instantiations, versus time $t$ for different values of $V$, $D=00$ and $w_t=40$ for all $t$, demonstrating the tradeoff in terms of $V$.  For large values of $t$, the utility achieved by the algorithm with $V=20$ is close to optimum while satisfying the constraints thereby confirming the optimality of the algorithm. It is important to note that the mixing coefficient can be easily estimated, and hence mixing condition can be verified through simulation.  

\section{Concluding Remarks}\label{sec:conl}
In this paper, we considered a distributed stochastic optimization problem with independent and asymptotically stationary states. We showed that this stochastic optimization problem is approximately equal to a \textbf{LP} that is a function of the limiting distribution of the state. For the proposed approximate DPP algorithm, we showed that with certain probabilities $\gamma_0$ and $\gamma_1$, the average cost and penalties are within constants of the optimal solution and the constraints, respectively, provided the waiting time $t > \text{a threshold}$. The threshold is in terms of the mixing coefficient that indicates the non-stationarity of the cost/penalties. The approximation errors capture the degree of non-stationarity (i.e., $\norm{\pi_t - \pi}_1$), the number of samples used to compute an estimate of the state distribution. Also, we have proved an almost sure convergence of the proposed algorithm to a constant close to the optimal. Finally, we presented simulations results to corroborate our theoretical findings. 
\appendices
\section{Mcdiarmid's Inequality} \label{app:Mcdiarmid_Inequality}
\begin{thrm} \label{thm:mcdiarmid}
\textbf{(See \cite{billingsley2013convergence})} Let $Z_1,Z_2,\ldots, Z_n$ be independent random variables all taking values in the set $\mathcal{Z}$. Let $f: \mathcal{Z}^n \rightarrow \mathbb{R}$ be a function that satisfies the following
\begin{equation}
\abs{f(z_1,\ldots,z_i,\ldots,z_n) - f(z_1,\ldots,z^{'}_i,\ldots,z_n)} \leq c_i, 
\end{equation}
for all $i=1,2,\ldots, n$. Then, for all $\epsilon > 0$
\begin{equation}
\Pr\{f - \mathbb{E} f > \epsilon\} \leq \exp\left\{ \frac{-2 \epsilon^2}{\sum_{i=1}^n c_i^2}\right\}.
\end{equation}
\end{thrm}

\section{Proof of Theorem \ref{thm:lppit_popt_relation}} \label{app:lppit_popt_relation}
In this proof, we use the fact that by decreasing the objective function and increasing the constraints $c_k$, $k=1,2,\ldots,K$ in $\mathbf{P_0}$ will result in a decreased optimal value. Consider the cost/penalties of the problem $\mathbf{P_0}$
\begin{eqnarray}
&&\hspace{-1.5cm}\limsup_{t \rightarrow \infty} \frac{1}{t}\sum_{\tau=0}^{t-1} \sum_{\omega \in \Omega} \pi_\tau(\omega) p_k(\alpha(\tau),\omega) \nonumber \\ 
&&\hspace{-0.5cm}\stackrel{(a)}{=}\limsup_{{t \rightarrow \infty,} {t > t^{'}}} \left[\frac{1}{t}\sum_{\tau=0}^{t-1} \sum_{\omega \in \Omega} \mathcal{P}_{i^{*}}(\omega) p_k(\tau) \right. \nonumber \\
&&+ \frac{1}{t}\sum_{\tau=0}^{t-1} \sum_{\omega \in \Omega} \left(\pi_\tau(\omega) -\pi(\omega) \right) p_k(\tau)  \nonumber\\
&&+ \left. \frac{1}{t}\sum_{\tau=0}^{t-1} \sum_{\omega \in \Omega} \left(\pi(\omega) -\mathcal{P}_{i^{*}}(\omega) \right) p_k(\tau)\right]  \label{eq:app1_costpenal_bound}
\end{eqnarray}
for $k = 0,1,2,\ldots,K$ and any $t^{'} > 0$. In the above, $p_k(\tau) \triangleq p_k(\alpha(\tau), \omega(\tau))$ and $(a)$ follows by adding and subtracting $\mathcal{P}_{i^{*}}(\omega)$ as mentioned earlier and $\pi(\omega)$. Since $\lim_{t \rightarrow \infty} \norm{\pi_t - \pi}_1 = 0$, for every $\nu > 0$, there exists a $t^{'} \in \mathbb{N}$ such that for all $t > t^{'}$, $\norm{\pi_t - \pi}_1 < \nu$. Using this $t^{'}$, and the fact that
\begin{eqnarray} 
&&\hspace{-1cm}\abs{\sum_{\omega \in \Omega} \left(\pi_\tau(\omega) -\pi(\omega) \right) p_k(t)} \leq \sum_{\omega \in \Omega} \abs{\left(\pi_t(\omega) -\pi(\omega) \right)} \abs{p_k(t)} \nonumber\\ 
&& \hspace{3.1cm}\leq \max\{\abs{p_{\text{max},k}}, \abs{p_{\text{min},k}}\}  \nu
\end{eqnarray}
for every $k$ and $t > t^{'}$, we have
\begin{equation} \label{eq:util_penal_term2}
-b_{\text{max},k} \nu \leq \frac{1}{t}\sum_{\tau=0}^{t-1} \sum_{\omega \in \Omega} \left(\pi_\tau(\omega) -\pi(\omega) \right) p_k(\tau)  \leq b_{\text{max},k} \nu,
\end{equation}
where $b_{\text{max},k} \triangleq \max\{\abs{p_{\text{max},k}}, \abs{p_{\text{min},k}}\}$. Similarly, we have 
\begin{eqnarray} \label{eq:util_penal_term3}
- b_{\text{max},k} d_{\pi,\mathcal{P}_{i^{*}}} \leq  \frac{1}{t}\sum_{\tau=0}^{t-1} \sum_{\omega \in \Omega} \left(\pi(\omega) -\mathcal{P}_{i^{*}}(\omega) \right) p_k(\tau) \nonumber \\
\leq b_{\text{max},k} d_{\pi,\mathcal{P}_{i^{*}}},
\end{eqnarray}
where $d_{\pi,\mathcal{P}_{i^{*}}}$ is as defined earlier.
Using \eqref{eq:util_penal_term2} and \eqref{eq:util_penal_term3} in \eqref{eq:app1_costpenal_bound}, we get the following lower bound for all $k=1,2,\ldots,K$.
\begin{eqnarray}
&&\limsup_{t \rightarrow \infty} \frac{1}{t}\sum_{\tau=0}^{t-1} \mathbb{E}p_k(\tau) \nonumber \\
&&\geq \limsup_{t \rightarrow \infty} \frac{1}{t}\sum_{\tau=0}^{t-1} \sum_{\omega \in \Omega} \mathcal{P}_{i^{*}}(\omega)p_k(\tau) - \Delta_{\pi,\mathcal{P}_{i^{*}}},
\end{eqnarray}
where $\Delta_{\pi,\mathcal{P}_{i^{*}}} = \max_{k=0,1,2,\ldots,K} b_{\text{max},k}  (d_{\pi,\mathcal{P}_{i^{*}}} + \nu)$. By using the above lower bound in $\mathbf{P_0}$, we get the following optimization problem 
\begin{eqnarray}
\mathbf{P_1}: &&\min_{\alpha(\tau) \in \mathcal{A}: \tau \in \mathbb{N}} \limsup_{t \rightarrow \infty} \frac{1}{t}\sum_{\tau=0}^{t-1} \mathbb{E} p_0(t) - \Delta_{\pi,\mathcal{P}_{i^{*}}} \nonumber \\
&&\hspace{-1.2cm}\text{s. t. } \limsup_{t \rightarrow \infty} \frac{1}{t}\sum_{\tau=0}^{t-1} \mathbb{E}p_k(t) \leq c_k + \Delta_{\pi,\mathcal{P}_{i^{*}}},~k=1,2,\ldots, K, \nonumber \\
&& \hspace{-0.8cm} \alpha_i(t) \text{ satisfies \eqref{eq:dist_condition}, } i=1,2,\ldots,N, \nonumber
\end{eqnarray} 
where the expectation is taken with respect to $\mathcal{P}_{i^{*}}$. Note that the optimal cost obtained by solving $\mathbf{P_1}$ is smaller than $p^{\text{opt}}$. Further, the term $\Delta_{\pi,\mathcal{P}_{i^{*}}}$ is independent of the control action. It is evident from $\mathbf{P_1}$ that it is equivalent to $\mathbf{P_0}$ where the states $\omega(t)$ is i.i.d. whose distribution is $\mathcal{P}_{i^{*}}$. Using Theorem $1$ of \cite{neely2016distributed}, it is easy to see that the solution to $\mathbf{P_1}$ is equal to $G(\Delta_{\pi,\mathcal{P}_{i^{*}}}) - \Delta_{\pi,\mathcal{P}_{i^{*}}}$,where $G(x)$ is as defined in \eqref{eq:gxlp}. 
Thus, from \textbf{Assumption 2}, we have that 
\begin{equation}
\abs{p^{\text{(pert)}}_{\mathcal{P}_{i^{*}}} - p^{(\text{opt})}_{\mathcal{P}_{i^{*}}}} < c \Delta_{\pi,\mathcal{P}_{i^{*}}} + \Delta_{\pi,\mathcal{P}_{i^{*}}}= (c+1) \Delta_{\pi,\mathcal{P}_{i^{*}}},
\end{equation}
where $p^{\text{(pert)}}_{\mathcal{P}_{i^{*}}}$ denotes the optimal cost of $\mathbf{P}_1$. This leads to $p^{\text{(pert)}}_{\mathcal{P}_{i^{*}}}  > p^{(\text{opt})}_{\mathcal{P}_{i^{*}}} - (c + 1) \Delta_{\pi,\mathcal{P}_{i^{*}}}$.
But, we know that $p^{\text{(pert)}}_{\mathcal{P}_{i^{*}}}\leq  p^{\text{(opt)}}$, which implies that
$p^{(\text{opt})}_{\mathcal{P}_{i^{*}}} < p^{\text{(opt)}} + (c + 1) \Delta_{\pi,\mathcal{P}_{i^{*}}}$.  $\blacksquare$
\section{Proof of Theorem \ref{thm:pacfirstresult}} \label{app:pacfirstresult}
Fix constants $\alpha_t \in \mathbb{N}$, $u_t$ and $v_t$ such that $u_t v_t = (t-\alpha_t)$, as shown in Fig. \ref{fig:time_div}. Let $\bar{p}_k(t) \triangleq \frac{1}{t}\sum_{\tau = 0}^{t-1} p_k(\tau)$. By adding and subtracting $\mathbb{E}\bar{p}_k(t)$ in the event in \eqref{eq:mcdiarmid_pac1}, we get
\begin{eqnarray}
&&\hspace{-0.7cm}\Pr\left\{\bar{p}_k(t) - \mathbb{E}{\bar{p}_k(t)} > \epsilon_{t,k}\right\} \nonumber\\
&&\hspace{-0.7cm}= \Pr\left\{ \frac{1}{{t}}\sum_{\tau=\alpha_t}^{t-1} (p_k(\tau) - \mathbb{E} p_k(\tau)) \right. \nonumber \\
&&\hspace{1cm} \left. > \epsilon_{t,k} - \frac{1}{{t}}\sum_{\tau=0}^{\alpha_t - 1} (p_k(\tau) - \mathbb{E} p_k(\tau)) \right\} \nonumber \\
&&\hspace{-0.7cm}\leq \Pr\left\{ \frac{1}{t - \alpha_t} \sum_{\tau=\alpha_t}^t (p_k(\tau) - \mathbb{E} p_k(\tau)) \right. \nonumber \\
&&\hspace{1cm}> \left. \frac{t \epsilon_{t,k} - {\alpha_t}(p_{\text{max},k} -  p_{\text{min},k})}{t-\alpha_t} \right\}, \label{eq:boundapp2_1}
\end{eqnarray}
where $\epsilon_{t,k} \triangleq \epsilon_k + c_k - \mathbb{E}{\bar{p}_k{(t)}}$, and the above inequality follows from the fact that $p_{\text{min},k} \leq p_k(t) \leq p_{\text{max},k}$ for all $t \in \mathbb{N}$. Note that we need $\epsilon_{t,k} > \frac{\alpha_t}{t}(p_{\text{max},k} -  p_{\text{min},k})$, which by using the definition of $\epsilon_{t,k}$ implies that $\epsilon_k  >  \mathbb{E}{\bar{p}_k{(t)}} - c_k +  \frac{\alpha_t}{t}(p_{\text{max},k} -  p_{\text{min},k})$. In order to apply the well known concentration inequalities, we need $p_k(t)$ to be independent across time $t$. Since $p_k(t)$'s are dependent across time, we use coupling argument to couple $p_k(t)$ process with an independent process $\tilde{p}_k(t)$ with the same distribution as $p_k(t)$. First, we divide the time slots from $\alpha_t$ to $t$ into $v_t$ blocks of size $u_t$ each. Thus, we need $t - \alpha_t = v_t u_t$. We divide the cost/penalties across blocks as $$\mathcal{S}_{i,k}^{(t)} \triangleq \{p_k(ju_t +i + \alpha_t): j = 0,1,2,\ldots, v_t -1\},$$ $i=0,1,2,\ldots,u_t-1$ and $k=0,1,\ldots,K$. See Fig. \ref{fig:time_div} for an illustration. Now, the time average of cost/penalties in \eqref{eq:boundapp2_1} can be written as 
\begin{equation}
\frac{1}{t - \alpha_t}\sum_{\tau = \alpha_t}^{t-1} p_k(\tau) = \frac{v_t}{t - \alpha_t}\sum_{i = 0}^{u_t-1} \Psi_{k,i,t}, 
\end{equation}
where $\Psi_{k,i,t} \triangleq \frac{1}{v_t}\sum_{s \in \mathcal{S}_{i,k}^{(t)}} p_k(s)$. Note that each term in $\Psi_{i,k,t}$ is at least $u_t$ slots apart. Using this, the above bound can be written as
\begin{eqnarray} 
&&\hspace{-1cm}\Pr\left\{\bar{p}_k(t) - \mathbb{E}{\bar{p}_k(t)} > \epsilon_{t,k}\right\} \nonumber \\ &\leq&\Pr\left\{\frac{v_t}{t}\sum_{i = 0}^{u_t-1} \Delta \Psi_{k,i,t} > \bar \epsilon_{t,k} \bigcap \mathcal{E}_{[\alpha_t:t]} \right\}  \nonumber \\
&&+ \Pr\left\{\frac{v_t}{t}\sum_{i = 0}^{u_t-1} \Delta \Psi_{k,i,t} > \bar \epsilon_{t,k} \bigcap \mathcal{E}_{[\alpha_t:t]}^c \right\}\nonumber \\
&\stackrel{(a)}{\leq}& \sum_{i=0}^{u_t - 1} \Pr\left\{\Delta \Psi_{k,i,t} > \bar \epsilon_{t,k} \left \vert \right. \mathcal{E}_{[\alpha_t:t]}^c \right\} + \sum_{\tau = \alpha_t}^t \Pr\left\{\mathcal{E}_{\delta,\tau}\right\}\nonumber \\  &\stackrel{(b)}{\leq}& \sum_{i=0}^{u_t - 1} \Pr\left\{\Delta \tilde{\Psi}_{k,i,t} > \bar \epsilon_{t,k}\left \vert \right. \mathcal{E}_{[\alpha_t:t]}^c\right\} + \sum_{\tau = \alpha_t}^t \Pr\left\{\mathcal{E}_{\delta,\tau}\right\} \nonumber \\ &&+ (t - \alpha_t) \beta_{\texttt{ALG},k}(u_t, \alpha_t \left \vert \right. \mathcal{E}_{[\alpha_t:t]}^c),  \label{eq:coupling_bound}
\end{eqnarray}
where $\bar \epsilon_{t,k} \triangleq \frac{t \epsilon_{t,k} - {\alpha_t}(p_{\text{max},k} -  p_{\text{min},k})}{t-\alpha_t}$, $\Delta \Psi_{k,i,t} \triangleq \Psi_{k,i,t} - \mathbb{E}{\Psi_{k,i,t}}$, $\tilde{\Psi}_{k,i,t} \triangleq \frac{1}{v_t}\sum_{\tau \in \mathcal{S}_{i,k}^{(t)}} \tilde{p}_k(\tau)$, and $\tilde{p}_k(\tau)$, conditioned on $\mathcal{E}_{[\alpha_t:t]}^c$, is an independent stochastic process having the same distribution as $p_k(\tau)$, $k=0,1,2,\ldots,K$. In the above, $(a)$ follows from the fact that the convex combination of the terms being greater than a constant implies that at least one of the term should be greater than the constant, and using the union bound. The inequality $(b)$ is obtained by applying proposition $1$ of \cite{kuznetsov2014generalization} to the indicator function $g \triangleq \mathbf{1}\{\Delta \Psi_{k,i,t} > \epsilon_{t,k}\}$ with expectation replaced by the conditional expectation $\mathbb{E}\left\{*\left \vert \right. \mathcal{E}_{[\alpha_t:t]}^c\right\}$. Recall that the event $\mathcal{E}_{\delta,\tau}$ corresponds to the error in decoding the correct index $i^*$ at time $\tau$ in \textbf{Step 1} of the algorithm, and $\mathcal{E}_{[\alpha_t:t]} = \bigcup_{\tau = \alpha_t}^t \mathcal{E}_{\delta,\tau}$. Note that $\Delta \tilde \Psi_{k,i,t}$ is the sum of $v_t$ independent random variables. Thus, by applying the Mcdiarmids inequality from Theorem \ref{thm:mcdiarmid} of Appendix \ref{app:Mcdiarmid_Inequality} along with the fact that $\tilde \Psi_{k,i,t} \leq (\Delta p)_{\max,k} = p_{\text{max},k} - p_{\text{min},k}$ for all $t \in \mathbb{N}$, we get 
\begin{equation}
\Pr\left\{\Delta \tilde \Psi_{k,i,t} > \epsilon_{t,k}\left \vert \right. \mathcal{E}_{[\alpha_t:t]}^c\right\} \leq \exp\left\{\frac{-2 \bar \epsilon_{t,k}^2 v_t^2}{((\Delta p)_{\text{max},k})^2} \right\}. 
\end{equation}

The above implies that 
\begin{equation}
\sum_{i=0}^{u_t-1} \Pr\left\{\Delta \tilde \Psi_{k,i,t} > \epsilon_{t,k}\left \vert \right. \mathcal{E}_{[\alpha_t:t]}^c\right\} \leq u_t \exp\left\{\frac{-2 \bar \epsilon_{t,k}^2 v_t^2}{((\Delta p)_{\text{max},k})^2} \right\}. 
\end{equation}

Using this in \eqref{eq:coupling_bound}, we get the desired result. $\blacksquare$

\section{Proof of Lemma \ref{lemm:prob_error}} \label{app:prob_error} 
First, we assume that at time $\tau$, $w_\tau$ samples are available. This requires $\tau - D - w_\tau +1 >0$, which implies that $\tau > D + w_\tau - 1$. Otherwise, the decoder will pick one of the $M_\delta$ measures uniformly at random resulting in a probability of error of $1/M_\delta$. This results in the second inequality in the theorem. Let $j_\tau$ be the output of the \textbf{Algorithm} at time slot $\tau$. In the following, we compute an upper bound on the probability of error when $\tau > D + w_\tau - 1$, i.e., 
\begin{eqnarray}
&&\hspace{-1cm}\Pr\{\mathcal{E}_{\delta,\tau}\} \nonumber \\
&=& \hspace{-0.4cm}\Pr\left\{ \bigcup_{j_\tau: j_\tau \neq i^*} \frac{1}{w_\tau} \sum_{s=\tau-D - w_\tau + 1}^{\tau-D} \log \mathcal{P}_{j_\tau}(\omega(s)) > f_{\tau,D,w_\tau}\right\} \nonumber\\
&\stackrel{(a)}{\leq}& \sum_{j_\tau:j_\tau \neq i^*} \Pr\left\{ \frac{1}{w_\tau} \sum_{s=\tau-D - w_\tau + 1}^{\tau-D} \log \left(\frac{\mathcal{P}_{j_\tau}(\omega(s))}{\mathcal{P}_{i^*}(\omega(s))}\right) > 0 \right\} \nonumber\\
&{\leq}& \sum_{j_\tau:j_\tau\neq i^*} \Pr\left\{ g_{\tau,D,w_\tau,j_\tau} - \mathcal{D}_{\tau,j_\tau} > -\mathcal{D}_{\tau,j_\tau} \right\},  
\end{eqnarray}
where $\mathcal{E}_{\delta,\tau}$ is the error in slot $\tau \in \mathbb{N}$ of step $1$ of the ADPP \textbf{Algorithm} due to incorrectly detecting the ``right distribution," $\mathcal{P}_{i^*} \in \mathcal{P}_c$ (see \eqref{eq:detect_covering_alg}). In the above, $g_{\tau,D,w_\tau,j_\tau}\triangleq\frac{1}{w_\tau} \sum_{s=\tau-D - w_\tau + 1}^{\tau-D} \log \left(\frac{\mathcal{P}_{j_\tau}(\omega(s))}{\mathcal{P}_{i^*}(\omega(s))}\right)$, $f_{\tau,D,w_\tau}\triangleq \frac{1}{w_\tau} \sum_{s=\tau-D - w_\tau + 1}^{\tau-D}  \log \mathcal{P}_{i^*}(\omega(s))$, and $\mathcal{D}_{\tau,j_\tau}$ is as defined in the Lemma. Note that $(a)$ follows from the union bound. By using the following boundedness property from \textbf{Assumption 3}, i.e.,  $\log \left(\frac{\mathcal{P}_{j_\tau}(\omega(s))}{\mathcal{P}_{i^*}(\omega(s))}\right)  \leq \log\left(\frac{\alpha_\delta}{\beta_\delta}\right)$, and using the Hoeffdings inequality (see \cite{billingsley2013convergence}), we get 
\begin{equation} \label{eq:peupperbound_stationary}
 \Pr\{\mathcal{E}_{\delta,\tau}\} \leq \exp\left\{- {2 \zeta_\delta \mathcal{D}_\tau^2 w_\tau} + \mathcal{H}(\mathcal{P},\delta) \right\},
\end{equation}
where $\zeta_\delta \triangleq {\left[\log\left(\frac{\alpha_\delta}{\beta_\delta}\right)\right]^2}$, $\mathcal{D}_\tau \triangleq \min_{j_\tau \neq i^*} \mathcal{D}_{\tau,j_\tau}$, and $\mathcal{H}(\mathcal{P},\delta) = \log M_\delta$ is the \emph{metric entropy}. Assuming $\alpha_t = \mathcal{O}(\sqrt{t})$ and $w_t = \mathcal{O}(\sqrt{t})$, we have that $\alpha_t > D + w_{\alpha_t} - 1 = D + t^{1/4} - 1$. This implies that $\Pr\{\mathcal{E}_{\delta,\tau}\} \preceq \exp\left\{- {2 \zeta_\delta \mathcal{D}_\tau^2 w_\tau} + \mathcal{H}(\mathcal{P},\delta) \right\}$, $\alpha_t \leq \tau \leq t-1$. Due to this, we have $\sum_{\tau=\alpha_t}^{t-1} \Pr\{\mathcal{E}_{\delta,\tau}\}  \preceq (t - \alpha_t) \exp\left\{- 2 \zeta_\delta \left[\min_{\alpha_t \leq \tau \leq t} \mathcal{D}_\tau\right]^2 N_{[\alpha_t:t]} + \mathcal{H}(\mathcal{P},\delta) \right\}
$, where $N_{[\alpha_t:t]} \triangleq \min_{\alpha_t \leq \tau \leq t} w_\tau$. This completes the proof. $\blacksquare$

\section{Proof of Theorem \ref{thm:mainresult1}} \label{app:mainresult1}
First, we upper bound $\mathbb{E}[\bar{p}_k(t)]$. As in \cite{neely2016distributed}, we consider the following instantaneous drift-plus-penalty expression denoted by $\mathcal{P}_{\tau,V} \triangleq \Delta(\tau+D) + V p_0(\tau)$
\begin{eqnarray}
\mathbb{E}\left[\mathcal{P}_{\tau,V}  \right] \hspace{-0.3cm} &=& \hspace{-0.3cm} \mathbb{E}\left[\mathcal{P}_{\tau,V}\left \vert \right.  \mathcal{E}_{\delta,\tau}^c \right] \Pr\{\mathcal{E}_{\delta,\tau}^c\} + \mathbb{E}\left[\mathcal{P}_{\tau,V}\left \vert \right. \mathcal{E}_{\delta,\tau} \right] \Pr\{\mathcal{E}_{\delta,\tau}\} \nonumber \\
&\leq& \hspace{-0.3cm}\mathbb{E}\left[\mathcal{P}_{\tau,V}\left \vert \right.  \mathcal{E}_{\delta,\tau}^c \right] + \mathbb{E}\left[\mathcal{P}_{\tau,V}\left \vert \right. \mathcal{E}_{\delta,\tau} \right] \Pr\{\mathcal{E}_{\delta,\tau}\}. \label{eq:dpp_bound}
\end{eqnarray}
where $\mathcal{E}_{\delta,\tau}$ is the error in slot $\tau \in \mathbb{N}$ of step $1$ of the ADPP \textbf{Algorithm} due to incorrectly detecting the ``right distribution," $\mathcal{P}_{i^*} \in \mathcal{P}_c$ (see \eqref{eq:detect_covering_alg}). 
Next, we will compute an upper bound on the second term in \eqref{eq:dpp_bound}, i.e., $\mathbb{E}\left[\mathcal{P}_{\tau,V}\left \vert \right. \mathcal{E}_{\delta,\tau} \right]$. Assume that the output of the \textbf{Algorithm} at time $\tau$ is $m_\tau^*$, and the corresponding induced probability be $\theta^{*}_m(\tau) = 1$ if $m = m_\tau^*$, zero otherwise. Now, we consider the following drift-plus-penalty bound on the second term in  \eqref{eq:dpp_bound} conditioned on $\mathbf{Q_\tau}$ 
\begin{eqnarray} 
\mathbb{E}\left[\mathcal{P}_{\tau,V}\left \vert \right. \mathcal{E}_{\delta,\tau}, \mathbf{Q_\tau} \right] &\stackrel{(a)}{\leq}& H_\tau + V \sum_{m=1}^F \theta^{*}_m(\tau) r_{0,\pi_\tau}^{(m)} \nonumber \\
&&+ \sum_{k=1}^K Q_k(\tau) \mathcal{C}_{k,\tau} \nonumber\\
&\stackrel{(b)}{\leq}& H_\tau + V p_{\text{max},0} + {\rho} \tau,
\end{eqnarray}
where $\rho \triangleq \sum_{k=1}^K (p_{\text{max},k} - c_k)^2$, $\mathcal{C}_{k,\tau} \triangleq \left[ \sum_{m=1}^F \theta^{*}_m(\tau) r_{k,\pi_\tau}^{(m)} - c_k\right]$, $H_\tau\triangleq (1 + 2D) B_\tau$, $B_\tau$ is as defined in \eqref{eq:btau}. In the above, $(a)$ follows from Lemma $6$ of \cite{neely2016distributed}, and $(b)$ follows from the fact that $Q_k(\tau) \leq \tau  (p_{\text{max},k} - c_k)$, $\mathcal{C}_{k,\tau} \leq (p_{\text{max},k} - c_k)$, and $p_{\text{max},k}$, $k=1,2,\ldots,K$ is the maximum cost/penalties. {Recall that $\mathbf{Q_\tau} = \{Q_1(\tau),Q_2(\tau),\ldots,Q_K(\tau)\}$ {and $Q_k(\tau+1) = \max\{Q_k(\tau)+p_k(\tau-D)-c_k,0\}, k=1,2,\ldots,K$. Since, $Q_k(0)=0$, $Q_k(1) = \max\{0+p_k(1-D)-c_k,0\} \leq (p_{\text{max},k}-c_k)$, and $Q_k(2)=\max\{Q_k(1)+p_k(2-D)-c_k,0\} \leq \max\{(p_{\text{max},k}-c_k)+p_k(2-D)-c_k,0\} \leq 2(p_{\text{max},k}-c_k)$. From the induction argument, we have $Q_k(\tau) \leq \tau  (p_{\text{max},k} - c_k)$, $k=1,2,\ldots,K$.}}. Taking the expectation of the above with respect to $\mathbf{Q_\tau}$ conditioned on $\mathcal{E}_{\delta, \tau}$ leads to the following result
\begin{equation}
\textbf{Result I:  } \mathbb{E}\left[\mathcal{P}_{\tau,V}\left \vert \right. \mathcal{E}_{\delta,\tau}\right] \leq H_\tau + V p_{\text{max},0} + \rho \tau.
\end{equation}
Applying Lemma $5$ of \cite{neely2016distributed} to the first term in \eqref{eq:dpp_bound} conditioned on $\mathbf{Q_\tau}$, we get 
\begin{eqnarray} 
&&\hspace{-1.0cm}\mathbb{E} \left[\mathcal{P}_{\tau,V} \left \vert \right. \mathbf{Q_\tau}, \mathcal{E}_{\delta,\tau}^c \right] \nonumber\\
&\leq& H_\tau + V \sum_{m=1}^F \theta^{*}_m(\tau) r_{0,\pi_\tau}^{(m)}  + \sum_{k=1}^K Q_k(\tau) \mathcal{C}_{k,\tau}\nonumber\\
&\leq&H_\tau + V \sum_{m=1}^F \theta^{*}_m(\tau) r_{0,\mathcal{P}_{i^*}}^{(m)} \nonumber\\
&+& V \sum_{m=1}^F \theta^{*}_m(\tau) \sum_{\omega \in \Omega} \Delta^{(\omega)}_{\pi,\mathcal{P}_{i^*}} p_0({\mathbf{S}^{m}}(\omega), \omega) \nonumber \\ &+& \hspace{-0.3cm} V \sum_{m=1}^F \theta^{*}_m(\tau) \sum_{\omega \in \Omega} \Delta^{(\omega)}_{\pi_\tau,\pi} p_0({\mathbf{S}^{m}}(\omega), \omega) + \sum_{k=1}^K Q_k(\tau) \mathcal{C}_{k,\tau}\nonumber \\
&\leq& \hspace{-0.1cm}H_\tau + V R_{0,\mathcal{P}_{i^*}}(\tau) + V  {J}^\pi_{\pi_\tau}  \nonumber\\
&+& \sum_{k=1}^K Q_k(\tau) \left[ R_k^{*}(\tau) - c_k\right], \label{eq:dpp_bound_appendix}
\end{eqnarray}
where $\mathcal{C}_{k,\tau} = \left[ \sum_{m=1}^F \theta^{*}_m(\tau) r_{k,\pi_\tau}^{(m)} - c_k\right]$ is as before, $R_{0,\mathcal{P}_{i^*}}(\tau) \triangleq\sum_{m=1}^F \theta_m^* r_{0,\mathcal{P}_{i^*}}^{(m)}$, $R_k^{*}(\tau) \triangleq \sum_{m=1}^F \theta_m^* r_{k,\pi_\tau}^{(m)}$, $\Delta^{(\omega)}_{\pi_\tau,\pi} \triangleq \abs{\pi_\tau(\omega) - \pi(\omega)}$, $H_\tau \triangleq B_\tau(1 + 2D)$, and ${J}^\pi_{\pi_\tau} \triangleq \max_{0\leq k \leq K} p_{\text{max},k} \left(\norm{\pi_\tau - \pi}_1 + \norm{\mathcal{P}_{i^*} - \pi}_1\right)$. Consider the following
\begin{eqnarray} 
&&\hspace{-2cm}\sum_{k=1}^K Q_k(\tau) \left[ R_k^{*}(\tau) - c_k\right]  \hspace{-0.3cm} \nonumber\\
&\leq&\sum_{k=1}^K Q_k(\tau) \left[\sum_{m=1}^F \theta^{*}_m(\tau) r_{k,\mathcal{P}_{i^*}}^{(m)} - c_k^{'} \right], \label{eq:temp1}
\end{eqnarray}
where $c_k^{'} \triangleq c_k - {J}^\pi_{\pi_\tau}$. We need $c_k > {J}^\pi_{\pi_\tau}$. The above inequality is obtained by (a) adding and subtracting $\mathcal{P}_{i^*}$, (b) using triangle inequality, and (c) following the steps that lead to the first three terms in \eqref{eq:dpp_bound_appendix}, and using the fact that $\norm{\mathcal{P}_{i^*} - \pi}_1 < \delta$. Substituting \eqref{eq:temp1} in \eqref{eq:dpp_bound_appendix}, we get 
\begin{eqnarray} 
&&\hspace{-1.5cm}\mathbb{E} \left[\mathcal{P}_{\tau,V} \left \vert \right. \mathbf{Q_\tau}, \mathcal{E}_{\delta,\tau}^c\right] \leq H_\tau +V  {J}^\pi_{\pi_\tau}  +V \sum_{m=1}^F \theta_m^* r_{0,\mathcal{P}_{i^*}}^{(m)} \nonumber\\
&+&  \sum_{k=1}^K Q_k(\tau)  \left[\sum_{m=1}^F \theta^{*}_m(\tau) r_{k,\mathcal{P}_{i^*}}^{(m)} - c_k^{'} \right]. \label{eq:dpp_alg}
\end{eqnarray}
Note that at each time slot $\tau$, the \textbf{Algorithm} chooses to minimize the right hand side of the above term when there is no error. Thus, choosing an alternative algorithm say $\theta_m$ will maximize the right hand side of \eqref{eq:dpp_alg}. Towards bounding the above further, let us choose a $\theta_m$ denoted $\theta_{m,\texttt{opt}}^{'}$ that optimally solves the problem $\mathbf{LP}_{\mathcal{P}_{i^*}}$ but with $c_k$ replaced by $c_k^{'}$. Further, let the corresponding optimal cost be $p^{'}_{\texttt{opt}}$. From \textbf{Assumption 2}, it follows that $p^{'}_{\texttt{opt}} < p^{(\text{opt})}_{\mathcal{P}_{i^{*}}} + c{J}^\pi_{\pi_\tau}$. Using the optimal  $\theta_{m,\texttt{opt}}^{'}$ in \eqref{eq:dpp_alg}, we get 
\begin{eqnarray}
 \mathbb{E} \left[\mathcal{P}_{\tau,V} \left \vert \right. \mathbf{Q_\tau}, \mathcal{E}_{\delta,\tau}^c\right] \hspace{-0.3cm}&\stackrel{(a)}{\leq}& \hspace{-0.3cm}V p^{'}_{\texttt{opt}} + H_\tau + V  {J}^\pi_{\pi_\tau} \nonumber\\
&<&\hspace{-0.3cm} V p^{(\text{opt})}_{\mathcal{P}_{i^{*}}} + V(c+1){J}^\pi_{\pi_\tau} + H_\tau,
\end{eqnarray}
where the inequality $(a)$ is obtained by noting that for $ \theta_m = \theta_{m,\texttt{opt}}^{'}$, $\left[\sum_{m=1}^F \theta_{m,\texttt{opt}}^{'} r_{k,\mathcal{P}_{i^*}}^{(m)} - c_k^{'} \right] < 0$, and $ p^{'}_{\texttt{opt}} = \sum_{m=1}^F \theta_{m,\texttt{opt}}^{'} r_{0,\mathcal{P}_{i^*}}^{(m)}$. Using $p^{(\text{opt})}_{\mathcal{P}_{i^{*}}} < p^{\text{(opt)}} + (c + 1) \Delta_{\pi,\mathcal{P}_{i^{*}}}$ from Theorem \ref{thm:lppit_popt_relation}, we get 
\begin{eqnarray} \label{eq:dpp_firstexpression}
{ \mathbb{E} \left[\mathcal{P}_{\tau,V} \left \vert \right. \mathbf{Q_\tau}, \mathcal{E}_{\delta,\tau}^c\right]}
\leq  V \mathcal{\psi_{\texttt{const}}} + {V(c+1){J}^\pi_{\pi_\tau} + H_\tau } ,
\end{eqnarray}
where $\mathcal{\psi_{\texttt{const}}}\triangleq p^{\text{(opt)}} + (c + 1) \Delta_{\pi,\mathcal{P}_{i^{*}}}$, and $\Delta_{\pi,\mathcal{P}_{i^{*}}}$ is as defined in Theorem \ref{thm:lppit_popt_relation}. Now, taking the expectation with respect to $\mathbf{Q_t}$ conditioned on $\mathcal{E}^c_{\delta,\tau}$, we get
\begin{equation}
 \textbf{Result II:  } { \mathbb{E} \left[\mathcal{P}_{\tau,V} \left \vert \right. \mathcal{E}_{\delta,\tau}^c\right]}
\leq  V \mathcal{\psi_{\texttt{const}}} + {V(c+1){J}^\pi_{\pi_\tau} + H_\tau }.
\end{equation}
Next, we borrow the results from Lemma \ref{lemm:prob_error} to obtain an upper bound (also, see \eqref{eq:peupperbound_stationary_lemma})  on the probability of error in  \eqref{eq:dpp_bound}, i.e.,  \\ \\
\textbf{Result III:  }
\begin{equation} \label{eq:app_peupperbound_stationary_lemma}
 \Pr\{\mathcal{E}_{\delta,\tau}\} \leq  P^{(\tau)}_{e,\texttt{up}} \triangleq \left\{ \begin{array}{cc} 
 q^{(\tau)}_{e,\texttt{up}} & \text{ if } \tau > D + w_\tau - 1,\\
 \frac{1}{M_\delta} & \text{otherwise},
\end{array}
\right.
\end{equation}
where $q^{(\tau)}_{e,\texttt{up}}$ is as defined in Lemma \ref{lemm:prob_error}. Using \textbf{Result I}, \textbf{Result II} and \textbf{Result III} in \eqref{eq:dpp_bound}, we get 
\begin{equation} \nonumber
\mathbb{E}\left[\mathcal{P}_{\tau,V}  \right] \leq V \mathcal{\psi_{\texttt{const}}} + {V(c+1){J}^\pi_{\pi_\tau} + H_\tau } + \mathcal{K}, 
\end{equation}
where $\mathcal{K} \triangleq  \left(H_\tau + V p_{\text{max},0}\right) P^{(\tau)}_{e,\texttt{up}}$. Summing the above over all slots $\tau = 0,1,2,\ldots, t-1$, and dividing by $t$, we get 
\begin{eqnarray} \label{eq:dpp_sumtau_bound}
&&\hspace{-1cm}\frac{\mathbb{E}\left[ \mathcal{L}(t+D) - \mathcal{L}(D)\right]}{t} + \frac{V}{t} \sum_{\tau=0}^{t-1} \mathbb{E} p_0(\tau) \nonumber \\
&\leq&\hspace{-0.3cm} V \psi_{\texttt{const}} + V(c+1) \bar{J}_t  \nonumber\\
&+& \hspace{-0.35cm}\bar{H}_t + \frac{(1+2D)}{t} \sum_{\tau = 0}^{t-1} B_{\tau} P_{e,\texttt{up}}^{(\tau)} + \frac{V p_{\text{max},0}}{t}\sum_{\tau=0}^{t-1} P_{e,\texttt{up}}^{(\tau)}.
\end{eqnarray}
Using the fact that $\mathcal{L}(t+D) \geq 0$, and $L(D) \leq C$ for some constant $C >0$, and after rearranging the terms, we get
\begin{eqnarray} \label{eq:diffpen_bound}
\mathbb{E} [\bar p_0(t)] - p^{\text{(opt)}}  &\leq& (c + 1) \Delta_{\pi,\mathcal{P}_{i^{*}}} + \psi_t(\delta),
\end{eqnarray}
where $\psi_t(\delta)$ is as defined in the theorem, and $\mathbb{E} [\bar p_0(t)] \triangleq \frac{1}{t}\sum_{\tau=0}^{t-1} \mathbb{E} p_0(\tau)$.
For any $\epsilon > 0$, choosing $\epsilon_0 = (c + 1) \Delta_{\pi,\mathcal{P}_{i^{*}}} + \psi_t(\delta) + \frac{{\alpha_t}{(p_{\text{max},k} -  p_{\text{min},k})}}{t-\alpha_t} + \epsilon$ satisfies the bound on $\epsilon_0$ in Theorem \ref{thm:pacfirstresult}. Again from Theorem \ref{thm:pacfirstresult} and the bound in \eqref{eq:diffpen_bound}, we have $\epsilon_{t,0}\triangleq \epsilon_0 + p^{(opt)} - \frac{1}{t}\sum_{\tau=0}^{t-1} \mathbb{E}p_0(\tau) \geq \epsilon$. Thus, using $\epsilon$ in place of $\epsilon_{t,0}$ in \eqref{eq:mcdiarmid_pac1}, $v_t = (t - \alpha_t)/u_t$, and substituting $\sum_{\tau=\alpha_t}^{t} P^{(\tau)}_{e,\texttt{up}} \leq (t - \alpha_t) S_{t,\delta}$ from Lemma \ref{lemm:prob_error},  we get the following upper bound 
\begin{eqnarray}
&&\hspace{-0.7cm}\Pr\left\{\frac{1}{t}\sum_{\tau=0}^{t-1} p_0(\tau) - p^{(opt)} >  \epsilon_0 \right\} \leq u_t \exp\left\{\frac{-2 \epsilon^2 (t - \alpha_t)^2}{(\Delta p)_{\text{max},0}^2 u_t^2}\right\} \nonumber\\
&&+ (t - \alpha_t) \left[\beta_{\texttt{ADPP},k}(u_t, \alpha_t \left \vert \right. \mathcal{E}_{\alpha_t, t}^c) \right. 
\left. +  S_{t,\delta} \right]
\end{eqnarray}
It is easy to verify that the above is less than or equal to $$\gamma_0 > (t - \alpha_t)\left[ \beta_{\texttt{ALG},k}(u_t,\alpha_t \left \vert \right. \mathcal{E}_{[\alpha_t:t]}) + S_{t,\delta} \right]$$ provided $t \in \mathcal{T}_{t,0}$, where $\mathcal{T}_{t,0}$ is as defined in the theorem. This proves the first part of the Theorem.

Multiplying \eqref{eq:dpp_sumtau_bound} by $t$, substituting for $\mathcal{L}(t+D)$, and using the fact that for all time slots $\tau$, there exists a constant $F$ such that $F  \geq  p^{\text{(opt)}} -  \mathbb{E} \left[ p_0(\tau) \right]$, we get (refer to the proof of Theorem $3$ of \cite{neely2016distributed} for details)
\begin{equation} \label{eq:queue_norm_bound}
\mathbb{E}\{\norm{\mathbf{Q}(t + D)}^2_2 \} \leq VFt + \Gamma_t,
\end{equation}
where $\Gamma_t$ is as defined in the theorem. Using Jensen's inequality, it follows from the above bound that
\begin{equation}
\frac{\mathbb{E}\{\abs{{Q_k}(t + D)} \}}{t} \leq Q_{\texttt{up}}(t) \triangleq \sqrt{\frac{VF}{t} + \frac{\Gamma_t}{ t^2}},
\end{equation}
for all $k=1,2,\ldots, K$. 
From Lemma $4$ of \cite{neely2016distributed}, we have $\mathbb{E}\{\bar{p}_k(t)\} \leq c_k + Q_{\texttt{up}}(t)$. 
Now, the right hand side of \eqref{eq:mcdiarmid_pac1} for $\epsilon_{t,k} = \epsilon$, $\epsilon_k = Q_{\texttt{up}}(t) + \epsilon$ is less than or equal to $\gamma_1$ provided $t \in \mathcal{T}_{t,1}$, where $\mathcal{T}_{t,1}$ is as defined in the theorem. 
 $\blacksquare$

\section{Proof of Theorem \ref{thm:d1_betabound}} \label{app:d1_betabound}
Note that conditioned on $\mathcal{E}_{[\alpha_t:t]}^c$, $\mathbf X_{t-s} \longrightarrow \mathbf{Q}_{t} \longrightarrow \mathbf X_{t}$ forms a Markov chain (see Fig. \ref{fig:graphicalmodel_1}). Thus, from\cite{polyanskiy2016strong}, we have
\begin{equation} \label{eq:mutual_bound_1}
I(\mathbf X_{t};\mathbf X_{t-s}\left \vert \right. \mathcal{E}_{[\alpha_t:t]}^c) \leq \eta_{{ch}_1} I(\mathbf{Q}_{t};\mathbf X_{t-s}\left \vert \right. \mathcal{E}_{[\alpha_t:t]}^c),
\end{equation}
where $\eta_{ch_1} \triangleq \sup_{q\neq q^{'}} \norm{\Pr\{\mathbf X_t \left \vert \right. \mathbf{Q}_t = q, \mathcal{E}_{[\alpha_t:t]}^c\} - \Pr\{\mathbf X_t \left \vert \right. \mathbf{Q}_t = q^{'}, \mathcal{E}_{[\alpha_t:t]}^c\}}_{\texttt{TV}}$ is the \emph{Dobrushin's} contraction coefficient for the channel $\mathbf{Q}_t$ to $\mathbf X_t$. By letting $\mathcal{Q}_t$ to denote the set of all possible vectors that $\mathbf{Q}_t$ can take, we have that $q, q^{'} \in \mathcal{Q}_t$.  Further, as $t$ increases, the cardinality of  $\mathcal{Q}_t$ grows. However, a ``small" change in the queue will not effect the strategy used, and hence $\eta_{ch_1} = 0$. We now make this observation more precise. As before, let $\mathcal{M}_t : \mathcal{Q}_t \rightarrow \{1,2,\ldots,F\}$ be the rule induced by the \texttt{ADPP} algorithm that determines the strategy given the queue at time $t$. Define $\mathbb{B}_t(m^*) \triangleq \{\mathbf{Q}_t \in \mathcal{Q}_t: \mathcal{M}_t(\mathbf{Q}_t) = m^*\}$. Note that $\mathbb{B}_t(m^*) \bigcap \mathbb{B}_t(m) = \phi$, $m \neq m^*$. Using the above set, we have the following equivalence relation. We say that $q \sim q^{'}$ if and only if $q, q^{'} \in \mathbb{B}_t(m^*)$ for some $m^* \in \{1,2,\ldots,F\}$, and $q \nsim q^{'}$ otherwise. It is easy to see that if $q \sim q^{'}$, then $\norm{\Pr\{\mathbf X_t \left \vert \right. \mathbf{Q}_t = q, \mathcal{E}_{[\alpha_t:t]}^c\} - \Pr\{\mathbf X_t \left \vert \right. \mathbf{Q}_t = q^{'}, \mathcal{E}_{[\alpha_t:t]}^c\}}_{\texttt{TV}} = 0$. Using this fact, we can equivalently write $\eta_{ch_1}$ as $\eta_{ch_1} = \sup_{q\nsim q^{'}} \norm{\Pr\{\mathbf X_t \left \vert \right. \mathbf{Q}_t = q, \mathcal{E}_{[\alpha_t:t]}^c\} - \Pr\{\mathbf X_t \left \vert \right. \mathbf{Q}_t = q^{'}, \mathcal{E}_{[\alpha_t:t]}^c\}}_{\texttt{TV}}$. When $q \nsim q^{'}$, we have $m \triangleq \mathcal{M}_t(q) \neq \mathcal{M}_t(q^{'}) \triangleq m^{'}$; this leads to the following 
\begin{eqnarray}
\eta_{ch_1} &\leq& \sup_{m \neq m^{'}} \norm{\Pr\{\mathbf X_t \left \vert \right. \mathcal{M}_t(\mathbf{Q}_t) = m, \mathcal{E}_{[\alpha_t:t]}^c\} \nonumber \\ 
	&&\hspace{1cm} - \Pr\{\mathbf X_t \left \vert \right. \mathcal{M}_t(\mathbf{Q}_t) = m^{'}, \mathcal{E}_{[\alpha_t:t]}^c\}}_{\texttt{TV}} \nonumber \\
&\leq& \max\left\{\frac{(e^\kappa - 1) }{2}, \frac{1}{2}\right\},  \label{eq:mutual_bound_kappa_1}
\end{eqnarray}
where the last inequality above follows from \textbf{Assumption 4}. Further, since $\kappa < \log 3$, we have $\frac{(e^\kappa - 1) }{2} < 1$. Using the fact that $\mathbf X_{t-s} \longrightarrow (\mathbf{Q}_{t-1}, \mathbf X_{t-1}) \longrightarrow \mathbf{Q}_t$ forms a Markov chain (see Fig. \ref{fig:graphicalmodel_1}), we can further bound the right hand side of \eqref{eq:mutual_bound_1} as follows
\begin{eqnarray}
\hspace{-0.6cm}I(\mathbf{Q}_{t};\mathbf X_{t-s}\left \vert \right. \mathcal{E}_{[\alpha_t:t]}^c) \hspace{-0.3cm} &\stackrel{(a)}{\leq}& \hspace{-0.3cm} I(\mathbf{Q}_{t-1}, \mathbf X_{t-1};\mathbf X_{t-s}\left \vert \right. \mathcal{E}_{[\alpha_t:t]}^c)\nonumber\\
&\stackrel{(b)}{\leq}& \hspace{-0.3cm} \eta_{{ch}_2} I(\mathbf{Q}_{t-2}, \mathbf X_{t-2};\mathbf X_{t-s}\left \vert \right. \mathcal{E}_{[\alpha_t:t]}^c), \label{mutual_bound_2}
\end{eqnarray}
where the \emph{Dobrushin's contraction coefficient} for the channel $(\mathbf{Q}_{t-1}, \mathbf X_{t-1})$ to $(\mathbf{Q}_{t-2}, \mathbf X_{t-2})$ conditioned on $\mathcal{E}_{[\alpha_t:t]}^c$ is given by
\begin{eqnarray}
\eta_{{ch}_2} \triangleq \sup_{(p,q)\neq (p^{'},q^{'})} \hspace{-0.5cm} && \norm{\Pr\{\mathbf X_{t-1}, \mathbf{Q}_{t-1} \left \vert \right. \mathcal{V}_{p,q,t}\} \nonumber \\
	&&- \Pr\{\mathbf X_{t-1}, \mathbf{Q}_{t-1} \left \vert \right. \mathcal{V}_{p^{'},q^{'},t}\}}_{\texttt{TV}}.
\end{eqnarray}  
In the above, $\mathcal{V}_{a,b,t} \triangleq \{\mathbf X_{t-2} = a, \mathbf{Q}_{t-2} = b, \mathcal{E}_{[\alpha_t:t]}^c\}$, $a \in \{p,p^{'}\}$ and $b \in \{q,q^{'}\}$. Further, $(a)$ and $(b)$ follow from the data processing inequality and SDPI for the channel $(\mathbf{Q}_{t-1}, \mathbf X_{t-1})$ to $(\mathbf{Q}_{t-2}, \mathbf X_{t-2})$, respectively \cite{polyanskiy2016strong}. Note that conditioned on $\mathcal{E}_{[\alpha_t:t]}^c$, the Dobrushin's contraction coefficients remain the same for all $(\mathbf{Q}_{t-i}, \mathbf X_{t-i})$ to $(\mathbf{Q}_{t-i-1}, \mathbf X_{t-i-1})$, $i=1,2,\ldots,s$. Using this argument and applying the SDPI repeatedly for \eqref{mutual_bound_2}, we get 
\begin{eqnarray}
I(\mathbf{Q}_{t};\mathbf X_{t-s}\left \vert \right. \mathcal{E}_{[\alpha_t:t]}^c) &\leq& \eta_{{ch}_2}^{s-2} I(\mathcal{Z}_{s,t};\mathbf X_{t-s}\left \vert \right. \mathcal{E}_{[\alpha_t:t]}^c)\nonumber\\
&& \hspace{-2.2cm}\leq \eta_{{ch}_2}^{s-2} \left(\log F + \log \abs{\Omega} + \log (K+1)\right), \label{eq:bound_iter_mutualinfo}
\end{eqnarray}
where $\mathcal{Z}_{s,t} \triangleq (\mathbf{Q}_{t-s+1}, \mathbf X_{t-s+1})$, and the last inequality follows from the fact that $I(\mathbf{Q}_{t-s+1}, \mathbf X_{t-s+1};\mathbf X_{t-s}\left \vert \right. \mathcal{E}_{[\alpha_t:t]}^c) \leq H(\mathbf X_{t-s} \left \vert \right.\mathcal{E}_{[\alpha_t:t]}^c) \leq H(\mathbf X_{t-s}) \leq \log(N_s)$, where $N_s = F \abs{\Omega} (K + 1)$ is the maximum number of possible values that $\mathbf X_\tau$ can take for all $\tau \in \mathbb{N}$. Using the  bound in \eqref{eq:bound_iter_mutualinfo}, and \eqref{eq:mutual_bound_kappa_1} in \eqref{eq:mutual_bound_1}, we get 
\begin{equation}
I(\mathbf X_{t};\mathbf X_{t-s}\left \vert \right. \mathcal{E}_{[\alpha_t:t]}^c) \leq \frac{\max\{(e^{\kappa}-1), 1\}\eta_{{ch}_2}^{s-2}}{2} \left[\log {N_s}\right]. \label{eq:bound_mutual_3}
\end{equation}
Now, it remains to bound $\eta_{{ch}_2}$. Towards this, consider 
\begin{eqnarray}
&&\hspace{-0.7cm}\eta_{{ch}_2} =\sup_{(p,q)\neq (p^{'},q^{'})} \frac{1}{2}\sum_{a,b}\left \vert\Pr\{\mathbf X_{t-1}=a, \mathbf{Q}_{t-1}=b \left \vert \right. \mathcal{V}_{p,q,t}\} \right. \nonumber \\ 
&&\left. ~~~~~ -\Pr\{\mathbf X_{t-1} = a, \mathbf{Q}_{t-1}=b \left \vert \right. \mathcal{V}_{p^{'},q^{'},t}\} \right \vert \nonumber\\
&&\hspace{-0.7cm}\stackrel{(a)}{=} \hspace{-0.5cm} \sup_{(p,q)\neq (p^{'},q^{'})} \frac{1}{2}\sum_{a,b}\left \vert  \Pr\{\mathbf{Q}_{t-1}=b \left \vert \right. \mathcal{V}_{p,q,t}\}  \right.\nonumber \\ 
&&\hspace{2.5cm}  \times \Pr\{\mathbf X_{t-1}=a \left \vert \right. \mathbf{Q}_{t-1}=b, \mathcal{E}_{[\alpha_t:t]}^c\} \nonumber\\ 
&&\hspace{1.5cm}-  \Pr\{\mathbf{Q}_{t-1}=b \left \vert \right. \mathcal{V}_{p^{'},q^{'},t}\} \nonumber \\ 
&&\hspace{2.5cm} \left. \times \Pr\{\mathbf X_{t-1}=a \left \vert \right. \mathbf{Q}_{t-1}=b, \mathcal{E}_{[\alpha_t:t]}^c\} \right \vert \nonumber\\
&&\hspace{-0.7cm}\stackrel{(b)}{=} \hspace{-0.5cm} \sup_{(p,q)\neq (p^{'},q^{'})} \frac{1}{2}\sum_{a,b}\left \vert \delta(b,\sigma_{p,q})\Pr\{\mathbf X_{t-1}=a \left \vert \right. \mathbf{Q}_{t-1}=b, \mathcal{E}_{[\alpha_t:t]}^c\} \right.\nonumber \\ &&\left.~~~- \delta(b,\sigma_{p^{'},q^{'}}) \Pr\{\mathbf X_{t-1}=a \left \vert \right. \mathbf{Q}_{t-1}=b, \mathcal{E}_{[\alpha_t:t]}^c\} \right \vert,
\end{eqnarray}
where $\mathcal{V}_{p,q,t}$ is as defined earlier, $\sigma_{x,y} \triangleq \max\{x + y -\mathbf C, 0\}$, $x \in \{p, p^{'}\}$, $y \in \{q,q^{'}\}$, and $\delta(.,.)$ is the Kroneckar delta function. In the above, $(a)$ is obtained by using the Bayes rule followed by the fact that $\mathbf X_{t-1}$ is independent of $\mathcal{V}_{p,q,t}$ and $\mathcal{V}_{p^{'},q^{'},t}$ conditioned on $\mathbf{Q}_{t-1}$ and $\mathcal{E}_{[\alpha_t:t]}^c$, and $(b)$ follows because $\mathbf{Q}_{t-1}$ is a deterministic function of $\mathbf{Q}_{t-2}$ and $\mathbf X_{\neq 0,t-2}$, i.e., $\mathbf{Q}_{t-1} = \max\{\mathbf{Q}_{t-2}+ \mathbf X_{t-2} - \mathbf C, 0\}$. Now, we have 
\begin{eqnarray}
\eta_{{ch}_2} \hspace{-0.2cm}&=&\hspace{-0.5cm} \sup_{(p,q)\neq (p^{'},q^{'})} \frac{1}{2}\sum_{a,b}\left \vert \Pr\{\mathbf X_{t-1}=a \left \vert \right. \mathbf{Q}_{t-1}=\sigma_{p,q}, \mathcal{E}_{[\alpha_t:t]}^c\} \right. \nonumber \\
&&\hspace{1.2cm}\left. -  \Pr\{\mathbf X_{t-1}=a \left \vert \right. \mathbf{Q}_{t-1}=\sigma_{p^{'},q^{'}}, \mathcal{E}_{[\alpha_t:t]}^c\} \right \vert \nonumber\\
&=&\hspace{-0.5cm} \sup_{(p,q)\neq (p^{'},q^{'})} \norm{\Pr\{\mathbf X_{t-1} \left \vert \right. \mathbf{Q}_{t-1}=\sigma_{p,q}, \mathcal{E}_{[\alpha_t:t]}^c\} \nonumber \\ 
	&& \hspace{1.2cm} -  \Pr\{\mathbf X_{t-1} \left \vert \right. \mathbf{Q}_{t-1}=\sigma_{p^{'},q^{'}}, \mathcal{E}_{[\alpha_t:t]}^c\} }_{\texttt{TV}} \nonumber\\
&\leq& \max\left\{\frac{e^{\kappa} - 1}{2}, \frac{1}{2}\right\} ,\label{eq:etakl_bound}
\end{eqnarray}  
where the last inequality above follows from the same argument that was used to obtain the bound on $\eta_{{ch_1}}$ in \eqref{eq:mutual_bound_kappa_1}. Using \eqref{eq:etakl_bound} in \eqref{eq:bound_mutual_3}, we get 
\begin{equation}
I(\mathbf X_{t};\mathbf X_{t-s}\left \vert \right. \mathcal{E}_{[\alpha_t:t]}^c) \leq \theta^{(s-1)} \left[\log {N_s}\right],
\end{equation}
where $\theta \triangleq \max\left\{\frac{(e^{\kappa}-1)}{2},\frac{1}{2}\right\}$. Substituting the above in \eqref{eq:betaone_bound_premitive}, we get the desired result in the Theorem. This completes the proof. $\blacksquare$

\section{Proof of Lemma \ref{lem:main1}} \label{app:as_D1_proof}
Since $u_t v_t = t-\alpha_t$, it is possible to choose $\alpha_t = \mathcal{O}(\sqrt t)$, ${u_t} = \mathcal{O}(\sqrt{t})$, and $v_t = \mathcal{O}(\sqrt t)$. Also, let $w_t = \mathcal{O}(\sqrt t)$, and thus, we have $$\frac{{\alpha_t}{(p_{\text{max},k} -  p_{\text{min},k})}}{t-\alpha_t} \rightarrow 0$$ as $t \rightarrow \infty$. Further,
 \begin{equation} \label{eq:app_main11}
(t -\sqrt{t}) \beta_{\texttt{ADPP},k}(u_t, \alpha_t \left \vert \right. \mathcal{E}_{[\alpha_t:t]}^c)  \preceq (t -\sqrt{t}){\frac{\theta^{{u_t}/2}}{\sqrt{2}} \left[\log \mu\right]} \rightarrow 0, 
\end{equation}
$k = 0,1,2,\ldots,K,$ as $t \rightarrow \infty$. Using $w_t = \mathcal{O}(\sqrt{t})$, and $\mathcal{H}(\mathcal{P},\delta) < \infty$ in the expression for $P_{e,\texttt{up}}^{(\tau)}$, we have 
\begin{eqnarray} \label{eq:app_main12}
&&\hspace{-0.7cm}\lim_{t \rightarrow \infty} P_{e,\texttt{up}}^{(t)} \nonumber\\
&&= \lim_{t \rightarrow \infty} (t- \sqrt{t}) \exp\left\{- {2 \zeta_\delta }(\min_{\alpha_t \leq \tau \leq t}\mathcal{D}_\tau)^2 w + \mathcal{H}(\mathcal{P},\delta) \right\} \nonumber \\
&&= 0.
\end{eqnarray}
By letting $V = \mathcal{O}(\sqrt{t})$ and the fact that $\sum_{\tau=0}^\infty P_{e,\texttt{up}}^{(\tau)} < \infty$, we have 
\begin{eqnarray} \label{eq:app_main13}
\lim_{t \rightarrow \infty} \psi_t(\delta) &=& \lim_{t \rightarrow \infty}\left[ \frac{\sqrt{t}(c+1){\bar{J}}_t + \bar{H}_t + C/t}{\sqrt{t}} \right. \nonumber \\ 
&&\left. + \frac{1+2D}{t\sqrt{t}}\sum_{\tau=0}^{t-1} B_\tau P_{e,\texttt{up}}^{(\tau)} + \frac{p_{\text{max},0}}{t} \sum_{\tau=0}^{t-1} P_{e,\texttt{up}}^{(\tau)}\right] \nonumber\\
&&= (c+1){\bar{J}}.
\end{eqnarray}
Since $F < \infty$, and $V=\mathcal{O}(\sqrt t)$, it follows that 
$\lim_{t \rightarrow \infty} Q_{\texttt{up}}(t)= \lim_{t \rightarrow \infty} \left[\sqrt{\frac{F}{\sqrt t} + \frac{\Gamma_t}{ t^2}}\right] = 0$.
Using this along with \eqref{eq:app_main13}, \eqref{eq:app_main11} and \eqref{eq:app_main12} in Theorem \ref{thm:mainresult1}, we get the desired result. $\blacksquare$

\section{Proof of Theorem \ref{thm:as_1}} \label{app:almost_sure_d1}
From the Borel-Cantelli Lemma, it follows that if 
\begin{equation}
\sum_{t=0}^\infty \Pr\left\{\frac{1}{t}\sum_{\tau=0}^{t-1} p_k(\tau) - c_k >  \epsilon_k \right\} < \infty,
\end{equation}
then $\lim_{t \rightarrow \infty} \frac{1}{t}\sum_{\tau=0}^{t-1} p_k(\tau) - c_k \leq  \epsilon_k $ almost surely \cite{billingsley2013convergence}. Using $u_t = \mathcal{O}(\sqrt t)$, $v_t = \mathcal{O}(\sqrt t)$, $\alpha_t = \mathcal{O}(\sqrt t)$, and an upper bound on the above from Theorem \ref{thm:pacfirstresult}, it suffices to show that 
\begin{eqnarray} \label{eq:app_as_split}
&&\hspace{-0.7cm}\sum_{t=0}^\infty \Pr\left\{\frac{1}{t}\sum_{\tau=0}^{t-1} p_k(\tau) - c_k >  \epsilon_k \right\} \nonumber \\
&&\leq \sum_{t=0}^\infty u_t \exp\left\{\frac{-2 \bar \epsilon_{t,k}^2 v_t^2}{((\Delta p)_{\text{max},k})^2}\right\}+ \sum_{t=0}^\infty\sum_{\tau = \alpha_t}^t \Pr\left\{\mathcal{E}_{\delta,\tau}\right\} \nonumber \\
&& +\sum_{t=0}^\infty (t - \alpha_{t}) \beta_{\texttt{ADPP},k}(u_t, \alpha_t \left \vert \right. \mathcal{E}_{[\alpha_t:t]}^c) < \infty, \label{eq:app_dev_alg}
\end{eqnarray}
where $\bar \epsilon_{t,k} = \frac{t \epsilon_{t,k} - {\alpha_t}(p_{\text{max},k} -  p_{\text{min},k})}{t-\alpha_t}$. Note that as $t \rightarrow \infty$, $\bar \epsilon_{t,k} \rightarrow \epsilon_{t,k}$. By choosing $u_t = \mathcal{O}(\sqrt{t})$ and $v_t = \mathcal{O}(\sqrt{t})$ as in the Theorem, it is easy to see that the first term above is finite since the summand is a product of a $\mathcal{O}({\sqrt{t}})$ term and an exponentially decreasing function of $t$. Using the result of Lemma \ref{lemm:prob_error} with $w_t = \mathcal{O}(\sqrt{t})$, it is easy to see that there exists a $t^* < \infty$ such that $t^* > D + w_{t^*} - 1$. Thus, for all $\tau < t^*$, $\Pr\{\mathcal{E}_{\delta,\tau}\} = 1/M_\delta$. Using this, the second term above can be written as
{
\begin{eqnarray}
&&\hspace{-0.7cm}\sum_{t=0}^{t^*}\sum_{\tau=\alpha_t}^{t-1} \Pr\{\mathcal{E}_{\delta,\tau}\} + \sum_{t=t^*+1}^{\infty}\sum_{\tau=\alpha_t}^{t-1} \Pr\{\mathcal{E}_{\delta,\tau}\} \nonumber\\
&&\hspace{-0.7cm} \leq \sum_{t=0}^{t^*} \frac{(t-\alpha_0)}{M_\delta} + \sum_{t=t^*+1}^{\infty}\sum_{\tau=\alpha_t}^{t-1} \Pr\{\mathcal{E}_{\delta,\tau}\} \nonumber \\
&&\hspace{-0.7cm} \leq \frac{(t^*-\alpha_0)t^*}{M_\delta} + \hspace{-0.3cm} \sum_{t=t^*+1}^{\infty} \hspace{-0.2cm} (t - \alpha_t) \exp\left\{- {\phi_{\tau,t,\delta}} + \mathcal{H}(\mathcal{P},\delta) \right\} 
\end{eqnarray}
where $\phi_{\tau,t,\delta} \triangleq 2 \zeta_\delta \left[\min_{\alpha_t \leq \tau \leq t} \mathcal{D}_\tau\right]^2 N_{[\alpha_t:t]}$, $N_{[\alpha_t:t]} \triangleq \min_{\alpha_t \leq \tau \leq t} w_\tau$ as in Lemma \ref{lemm:prob_error}. Since $t^*$ is finite, we can say that the first term in the second inequality is bounded. Since $\alpha_t = \mathcal{O}(\sqrt{t})$, $(t - \alpha_t) \exp\left\{- {\phi_{\tau,t,\delta}} + \mathcal{H}(\mathcal{P},\delta) \right\} \rightarrow 0$ exponentially fast as $t \rightarrow \infty$. Hence, the second inequality is also bounded.} Now, using the result from \eqref{eq:mixing_d1} of Corollary \ref{corr:mixing_d1}, the third term in \eqref{eq:app_dev_alg} becomes   
\begin{equation}
\sum_{t=0}^\infty (t - \alpha_{t}) \beta_{\texttt{ADPP},k}(u_t, \alpha_t \left \vert \right. \mathcal{E}_{[\alpha_t:t]}^c) \doteq \frac{\left[\log \mu\right]}{\sqrt{2}}\sum_{t=0}^\infty (t - \sqrt{t}) {\theta^{\sqrt{t}/2}},
\end{equation}
which is finite since $\theta < 1$ (see Corollary \ref{corr:mixing_d1}). Thus, we have $p_k(\tau) - c_k \leq  \epsilon_k $ almost surely for all $k$. Next, we need to show that the first term in \eqref{eq:app_as_split} is finite. Towards this it suffices to show that $\bar{\epsilon}_{t,k} < \infty$ since $u_t =\mathcal{O}(\sqrt t)$ and $v_t =\mathcal{O}(\sqrt t)$. Equivalently, from the definition of $\bar{\epsilon}_{t,k}$ in Theorem \ref{thm:pacfirstresult}, we need to show that $\lim_{t \rightarrow \infty} {\epsilon}_{t,k} < \infty$. From the proof of Theorem \ref{thm:mainresult1}, using $\infty > \epsilon_0 > \frac{1}{t}\sum_{\tau=0}^{t-1} \mathbb{E}p_0(\tau) - p^{\text{(opt)}} + \frac{{\alpha_t}{(p_{0,\texttt{max}} -  p_{0,\texttt{min}})}}{t-\alpha_t} > (c + 1) \Delta_{\pi,\mathcal{P}_{i^{*}}} + \psi_t(\delta) + \frac{{\alpha_t}{(p_{0,\texttt{max}} -  p_{0,\texttt{min}})}}{t-\alpha_t} + \epsilon$ and $\epsilon_k = Q_{\texttt{up}}(t) + \epsilon$, we get the desired result. $\blacksquare$

\section{Proof of Lemma \ref{lemm:etach1_bound}} \label{app:etach1_bound}
Consider the following expression from the definition of $\eta_{\texttt{ch}_1}$ in \eqref{eq:defn_etach1} with $\gamma \triangleq \{\gamma_{D},\gamma_{D-2},\ldots, \gamma_1\}$.
\begin{eqnarray}
&&\hspace{-0.7cm}\Pr\left\{{\mathbf{X}_{0,t}} \left \vert \right.{\mathbf{Q}_{0,t}} = \gamma \right\} \nonumber \\
&&\hspace{-0.7cm}= \Pr\{\mathbf {X}_{t} \left \vert \right. \mathbf Q_t = \gamma_D\} \times \Pr\{\mathbf{X}_{t-1} \left \vert \right. \mathcal{B}_1\} \times \ldots \nonumber\\ 
&&\hspace{-0.7cm} \ldots \times \Pr\{\mathbf {X}_{t-D + 1} \left \vert \right. \mathcal{B}_2\} \nonumber\\
&&\hspace{-0.7cm}= \Pr\{\mathbf {X}_{t} \left \vert \right. \mathbf Q_t = \gamma_D\} \times \Pr\{\mathbf {X}_{t-1} \left \vert \right. \mathbf Q_{t-1}= \gamma_{D-1}\} \times \ldots \times \nonumber\\&& \Pr\{\mathbf {X}_{t-D + 1} \left \vert \right. \mathbf Q_{t-D+1} = \gamma_1\},\label{eq:eta_ch1_prod}
\end{eqnarray}
where $\mathcal{B}_1 \triangleq \{\mathbf Q_t= \gamma_D, \mathbf Q_{t-1}= \gamma_{D-1}\}$, $\mathcal{B}_2 \triangleq \{\mathbf Q_t= \gamma_D, \mathbf Q_{t-1}= \gamma_{D-1},\ldots,\mathbf Q_{t-D+1} = \gamma_1\}$, and the last step follows from the Markov chain property, i.e., conditioned on $\mathbf Q_{t-i}$, $\mathbf {X}_{t-i}$ is independent of $\mathbf Q_{t-i+j}$, $j = 0,1,\ldots,i$. As in the proof of Theorem \ref{thm:d1_betabound}, we let $\mathcal{Q}_{t}$ to denote the set of all possible vectors that $\mathbf {Q}_{t}$ can take.  Further, as $t$ increases, the cardinality of  $\mathbf {Q}_{t}$ grows. However, a ``small" change in the queue will not effect the strategy used, and hence $\eta_{\texttt{ch}_1} = 0$. We now make this observation more precise. Let $\mathcal{M}_t : \mathcal{Q}_t \rightarrow \{1,2,\ldots,F\}$ be the rule induced by the \texttt{ADPP} algorithm that determines the strategy given the queue at time $t$. For any strategy $m^*$, define $\mathbb{B}_t(m^*) \triangleq \{\mathbf{Q}_t \in \mathcal{Q}_t: \mathcal{M}_t(\mathbf Q_t) = m^*\}$. Using the above set, we have the following equivalence relation. We say that $q \sim q^{'}$ if and only if $q, q^{'} \in \mathbb{B}_t(m^*)$ for some $m^* \in \{1,2,\ldots,F\}$, and $q \nsim q^{'}$ otherwise. It is easy to see that if $q \sim q^{'}$, then $\norm{\Pr\{\mathbf X_{t-i} \left \vert \right.\mathbf  {Q}_{t-i} = q, \mathcal{E}_{[\alpha_t:t]}^c\} - \Pr\{\mathbf X_{t-i} \left \vert \right. \mathbf {Q}_{t-i} = q^{'}, \mathcal{E}_{[\alpha_t:t]}^c\}}_{\texttt{TV}} = 0$, $i=0,1,\ldots,D-1$. Using this fact, we can equivalently write $\eta_{\texttt{ch}_1}$ as $\eta_{\texttt{ch}_1} = \sup_{q\nsim q^{'}} \norm{\Pr\{\mathbf X_t \left \vert \right. \mathbf {Q}_t = q, \mathcal{E}_{[\alpha_t:t]}^c\} - \Pr\{\mathbf X_t \left \vert \right. \mathbf {Q}_t = q^{'}, \mathcal{E}_{[\alpha_t:t]}^c\}}_{\texttt{TV}}$. However, when $q \nsim q^{'}$, we have $\mathcal{M}_t(q) \neq \mathcal{M}_t(q^{'})$; using this and \eqref{eq:eta_ch1_prod} in the definition of $\eta_{\texttt{ch}_1}$ leads to the following  
\begin{eqnarray}
\eta_{\texttt{ch}_1} \hspace{-0.4cm}&=& \hspace{-0.4cm} \sup_{\gamma \neq \gamma^{'}} \left \vert \left \vert{\prod_{k=0}^{D-1}\Pr\left\{{X}_{t-k} \left \vert \right.{Q}_{t-k} = \gamma_{D-k} \right\}} \right. \right.\nonumber \\ &&\left. \left. {-\prod_{k=0}^{D-1}\Pr\left\{{X}_{t-k} \left \vert \right.{Q}_{t-k} = \gamma^{'}_{D-k} \right\}}\right \vert \right\vert_{\texttt{TV}} \nonumber \\
&\stackrel{(a)}{=}&\hspace{-0.4cm} \sup_{m \neq m^{'}} \left \vert \left \vert \prod_{k=0}^{D-1} \Pr\{X_{t-k} \left \vert \right. \mathcal{M}_{t-k}({Q}_{t-k}) = m, \mathcal{E}_{[\alpha_t:t]}^c\} \right. \right. \nonumber \\ 
&&\left. \left. - \prod_{k=0}^{D-1}\Pr\{X_{t-k} \left \vert \right. \mathcal{M}_{t-k}({Q}_{t-k}) = m^{'}, \mathcal{E}_{[\alpha_t:t]}^c\}\right \vert \right \vert_{\texttt{TV}} \nonumber\\
&{=}& \hspace{-0.4cm}\sup_{\gamma \neq \gamma^{'}}  \left \vert \left \vert\prod_{k=0}^{D-1}\Pr\{X_{t-k} \left \vert \right. \mathcal{M}_{t-k}({Q}_{t-k}) = m^{'}, \mathcal{E}_{[\alpha_t:t]}^c\} \right. \right. \nonumber \\
&&\hspace{-0.45cm}\left. \left. \left(\prod_{k=0}^{D-1}\frac{\Pr\{X_{t-k} \left \vert \right. \mathcal{M}_{t-k}({Q}_{t-k}) = m, \mathcal{E}_{[\alpha_t:t]}^c\}}{\Pr\{X_{t-k} \left \vert \right. \mathcal{M}_{t-k}({Q}_{t-k}) = m^{'}, \mathcal{E}_{[\alpha_t:t]}^c\}} - 1\right)\right \vert \right\vert_{\texttt{TV}}   \nonumber \\
&\stackrel{(b)}{\leq}& \hspace{-0.2cm}\max\left\{\frac{(\exp\{\kappa D\} -1)}{2}, \frac{1}{2}\right\} < 1, \label{eq:eta_ch1_bound1}
\end{eqnarray}
where $(a)$ follows by substituting \eqref{eq:eta_ch1_prod} in the definition of $\eta_{\texttt{ch}_1}$, and $(b)$ follows from the \textbf{Assumption $4$} and the definition of the total variational norm. This completes the proof. $\blacksquare$

\section{Proof of Lemma \ref{lemm:etachj_bound}} \label{app:etachj_bound}
Note that $\mathbf{Q}_{j-1,t}$ is a deterministic function of $\mathbf{X}_{j,t}$ and $\mathbf{Q}_{j,t}$, i.e., $\mathbf{Q}_{j-1,t} = \max \left\{\mathbf{Q}_{j,t} + \mathbf{X}_{j,t} - \mathbf C,0\right\}$,
where $\mathbf C \triangleq \{\underbrace{c_1,\ldots,c_1}_{{D} \text{ times}},\underbrace{c_2,\ldots,c_2}_{{D} \text{ times}},\ldots,\underbrace{c_K,\ldots,c_K}_{{D} \text{ times}} \}$. Using \eqref{eq:eta_chj_equality}, $\eta_{ch_j}$ can be written as
\begin{eqnarray}
&&\hspace{-0.7cm}\eta_{\texttt{ch}_j} =\hspace{-0.4cm} \sup_{(p,q) \neq (p^{'}, q^{'})} \left \vert \left \vert  \Pr\left\{ {\mathbf{Q}_{j-1,t}=b} \left \vert \right. {\mathbf{X}_{j,t}} = p, {\mathbf{Q}_{j,t}} = q, {\mathcal{E}_{[\alpha_t:t]}^c} \right\} \right. \right.  \nonumber \\ 
&& \hspace{1.5cm} \times  \Pr\left\{{\mathbf{X}_{j-1,t}=a} \left \vert \right. {\mathbf{Q}_{j-1,t}} = b,{\mathcal{E}_{[\alpha_t:t]}^c} \right\}  \nonumber \\ 
&& \hspace{0.7cm} - \Pr\left\{{\mathbf{Q}_{j-1,t}=b} \left \vert \right.{\mathbf{X}_{j,t}} = p^{'}, {\mathbf{Q}_{j,t}} = q^{'}, {\mathcal{E}_{[\alpha_t:t]}^c} \right\} \nonumber \\
&&\hspace{1.5cm}\left. \left.\times \Pr\left\{{\mathbf{X}_{j-1,t}=a} \left \vert \right. {\mathbf{Q}_{j-1,t}} = b, {\mathcal{E}_{[\alpha_t:t]}^c} \right\}  \right \vert \right \vert_{\texttt{TV}} \nonumber \\
&&\hspace{-0.7cm} \stackrel{(a)}{=} \hspace{-0.3cm} \sup_{(p,q) \neq (p^{'}, q^{'})} \left \vert \left \vert \delta(b,\sigma_{p,q}) \Pr\left\{{\mathbf{X}_{j-1,t}=a} \left \vert \right. {\mathbf{Q}_{j-1,t}} = b,{\mathcal{E}_{[\alpha_t:t]}^c} \right\} \right. \right. \nonumber \\
&& \left. \left. -\delta(b,\sigma_{p',q'}) \Pr\left\{{\mathbf{X}_{j-1,t}=a} \left \vert \right. {\mathbf{Q}_{j-1,t}} = b,{\mathcal{E}_{[\alpha_t:t]}^c} \right\} \right \vert \right \vert_{\texttt{TV}} \nonumber \\
&&\hspace{-0.7cm} \leq \sup_{(p,q) \neq (p^{'}, q^{'})} \left \vert \left \vert \Pr\left\{{\mathbf{X}_{j-1,t}=a} \left \vert \right. {\mathbf{Q}_{j-1,t}} = \sigma_{p,q},{\mathcal{E}_{[\alpha_t:t]}^c} \right\} \right. \right. \nonumber \\
&&\hspace{0cm} \left. \left. - \Pr\left\{{\mathbf{X}_{j-1,t}=a} \left \vert \right. {\mathbf{Q}_{j-1,t}} = \sigma_{p',q'},{\mathcal{E}_{[\alpha_t:t]}^c} \right\} \right \vert \right \vert_{\texttt{TV}} \nonumber \\
&&\hspace{-0.7cm}< \max\left\{\frac{\exp\{\kappa D\} - 1}{2}, \frac{1}{2}\right\} < 1, 
\end{eqnarray}
where $\sigma_{x,y} \triangleq \max\{x + y -\mathbf C, 0\}$, $x \in \{p, p^{'}\}$, $y \in \{q,q^{'}\}$, and $\delta(.,.)$ is the Kroneckar delta function as in Theorem \ref{thm:d1_betabound} and $(a)$ follows from the fact that $\mathbf{Q}_{j-1,t}$ is a deterministic function of $\mathbf{X}_{j,t}$ and $\mathbf{Q}_{j,t}$, i.e., $\mathbf{Q}_{j-1,t} = \max \left\{\mathbf{Q}_{j,t} + \mathbf{X}_{j,t} - \mathbf C,0\right\}$. The last inequality follows from the argument used in the proof of Lemma \ref{lemm:etach1_bound}. $\blacksquare$

\bibliographystyle{IEEEtran}
\bibliography{IEEEabrv,DPP2016}

\end{document}